\documentclass{article}

\usepackage{microtype}
\usepackage{graphicx}
\usepackage{subfigure}
\usepackage{booktabs} 
\usepackage{multirow}

\usepackage[normalem]{ulem} 
\usepackage{xcolor}

\usepackage{hyperref}

\usepackage[preprint]{icml2026}

\usepackage{amsmath}
\usepackage{amssymb}
\usepackage{mathtools}
\usepackage{amsthm}

\usepackage[capitalize,noabbrev]{cleveref}

\usepackage[most]{tcolorbox}
\usepackage{listings}

\usepackage{float}
\definecolor{keywordcolor}{rgb}{0.7, 0.1, 0.1}   
\definecolor{commentcolor}{rgb}{0.4, 0.4, 0.4}   
\definecolor{stringcolor}{rgb}{0.5, 0.3, 0.2}    
\definecolor{symbolcolor}{rgb}{0.1, 0.2, 0.7}    
\definecolor{sortcolor}{rgb}{0.1, 0.5, 0.1}      
\definecolor{attributecolor}{rgb}{0.7, 0.1, 0.1} 
\definecolor{errorcolor}{rgb}{1, 0, 0}

\definecolor{jsonkeyword}{rgb}{0.5, 0, 0.5}      
\definecolor{jsonstring}{rgb}{0.2, 0.5, 0.2}     
\definecolor{jsoncomment}{rgb}{0.4, 0.4, 0.4}    
\definecolor{jsonsymbol}{rgb}{0.6, 0.1, 0.1}     
\definecolor{jsonnumber}{rgb}{0.1, 0.1, 0.6}     

\newcommand{\redwave}{
  \bgroup
  \markoverwith{\textcolor{red}{\uwave{\phantom{X}}}}\ULon
}

\theoremstyle{plain}
\newtheorem{theorem}{Theorem}[section]

\theoremstyle{definition}
\newtheorem{definition}[theorem]{Definition}

\theoremstyle{remark}

\usepackage[textsize=tiny]{todonotes}

\icmltitlerunning{Construction–Verification: A Benchmark for Applied Mathematics in Lean 4}
\begin{document}

\twocolumn[
\icmltitle{Construction–Verification: A Benchmark for  Formalizing \\ Applied Mathematics in Lean 4}

\icmlsetsymbol{equal}{*}

\begin{icmlauthorlist}
\icmlauthor{Bowen Yang}{sms}
\icmlauthor{Yi Yuan}{sms}
\icmlauthor{Chenyi Li}{sms}
\icmlauthor{Ziyu Wang}{sms}
\icmlauthor{Liangqi Li}{ant}
\icmlauthor{Bo Zhang}{ant}
\icmlauthor{Zhe Li}{ant}
\icmlauthor{Zaiwen Wen}{sms}
\end{icmlauthorlist}

\icmlaffiliation{sms}{Peking University, Beijing, China}
\icmlaffiliation{ant}{Ant Digital Technologies, Ant Group}

\icmlcorrespondingauthor{Zaiwen Wen}{wenzw@pku.edu.cn}

\icmlkeywords{Machine Learning, ICML}

\vskip 0.3in
]

\printAffiliationsAndNotice{}

\begin{abstract}
Recent advances in large language models have demonstrated impressive capabilities in mathematical formalization. However, existing benchmarks focus on logical verification of declarative propositions, often neglecting the task of explicitly synthesizing solutions. This limitation is particularly acute in applied mathematics domains, where the goal is frequently to derive concrete values or executable algorithms rather than solely proving theorems. To address this, we introduce a Lean 4 framework that enforces a construction-verification workflow, compelling the agent to define explicit solutions before proving their correctness. We curate a comprehensive benchmark \textbf{AMBER} (\textbf{A}pplied \textbf{M}athematics \textbf{BE}nchmark for \textbf{R}easoning) spanning core domains of applied mathematics, including convex analysis, optimization, numerical algebra, and high-dimensional probability. Aside from theorem proving, our benchmark features complex tasks such as evaluation, algorithm design, and representation transformation. Experiments reveal that current models face significant difficulties with these constructive tasks. Notably, we observe that general-purpose reasoning models consistently outperform specialized theorem provers. We attribute this to a degradation of instruction following capabilities in specialized models. Fine-tuning on proof corpora appears to induce ``tactical overfitting", compromising the ability to adhere to complex constructive requirements, whereas general models retain the versatility needed for multi-task formal reasoning.
\end{abstract}

\section{Introduction}
\label{sec:introduction}
The growing complexity of modern mathematics, coupled with the potentially severe consequences of errors in mathematical proofs, has created an urgent need for rapid verification and error detection. This demand underscores the critical importance of formal mathematics. Interactive theorem provers (ITP), which utilize formal mathematical languages, such as HOL Light \cite{harrison1996hol}, Coq \cite{huet1997coq}, Isabelle \cite{paulson1994isabelle}, and Lean \cite{delean,moura2021lean}, address this challenge by enabling the automated verification of proofs with absolute certainty, significantly mitigating the risk of human error inherent in traditional manual review.

However, this assurance of correctness comes at the cost of a significant usability barrier, as constructing formal proofs demands expert knowledge of extensive libraries and is often a painstakingly manual process due to repetitive tasks. This challenge has catalyzed a promising research direction: leveraging large language models (LLMs) to automate reasoning within these formal systems. Notably, both general-purpose LLMs (e.g., DeepSeek-V3.2-Thinking \cite{liu2025deepseek}, Kimi K2 \cite{team2025kimi}, Gemini-3 Pro, and fine-tuned models, such as Seed-Prover \cite{chen2025seed}, DeepSeek-Prover-V2 \cite{ren2025deepseek}, have demonstrated significant success in automated formalization, particularly in proving high school contest problems.

To evaluate and improve the capabilities of large language models in automated formalization, researchers have developed a variety of specialized benchmarks, such as MiniF2F \cite{zheng2021minif2f}, ProofNet \cite{azerbayev2023proofnet}. However, existing Lean-based benchmarks primarily focus on elementary or pure mathematics, with notably limited coverage of undergraduate-level or higher-level applied mathematics. More critically, these benchmarks provide a systematically inadequate representation of core concepts in applied mathematics, such as numerical evaluation, algorithm design and representation transformation, which ultimately hinders the expansion of the formalized mathematical knowledge frontier.

To bridge this gap, we propose that effectively evaluating applied mathematical reasoning requires a paradigm shift from verifying existential propositions to synthesizing constructive solutions. Unlike pure theorem proving, where establishing the existence of a solution often suffices, applied mathematics necessitates the derivation of explicit values and executable algorithms. This distinction demands a rigorous formalization approach that structurally enforces the ``construction-verification" workflow, preventing agents from exploiting non-constructive shortcuts. By aligning formal tasks with the computational intuition of fields like convex optimization and numerical algebra, we aim to establish a more faithful standard for automated reasoning in higher-level mathematics.

Furthermore, evaluating current models within this constructive framework reveals a fundamental trade-off: intensive specialization in formal logic often diminishes the general reasoning and instruction-following skills necessary for complex mathematical modeling. We observe that the prevailing strategy of fine-tuning on proof corpora, while effective for standard deduction, may inadvertently compromise the instruction-following versatility required for multi-stage mathematical modeling. This phenomenon highlights a significant trade-off in current neuro-symbolic systems: the tension between local tactical proficiency and the high-level algorithmic synthesis essential for applied mathematics.

In summary, our contributions are as follows.

\begin{itemize}
    \item We develop specialized formalization paradigms tailored for distinct categories of applied mathematics. Specifically, we design a construction-verification pattern for evaluation problems, algorithm design and representation transformations.
    \item We construct an \textbf{AMBER} (\textbf{A}pplied \textbf{M}athematics \textbf{BE}nchmark for \textbf{R}easoning) benchmark covering core applied domains (convex analysis, optimization, numerical algebra, probability) with varying difficulty levels, from standard undergraduate exercises to PhD-level research problems. 
    \item We conduct comprehensive evaluations demonstrating that current models struggle with constructive applied mathematics. Our analysis highlights the trade-off between proof-search specialization and general constructive capability, offering new insights for future neuro-symbolic research. 
\end{itemize}

\section{Related Works}
\label{sec:problem_related_works}

\textbf{Formal Benchmarks. } Formal benchmarks play a crucial role in enabling rigorous and reproducible assessment of automated theorem proving capabilities. In the domain of pre-university mathematics, the MiniF2F benchmark \cite{zheng2021minif2f} focuses on problems drawn from mathematical competitions, a direction also explored by FIMO \cite{liu2023fimo} and PutnamBench \cite{tsoukalas2024putnambench}. At the university level, ProofNet \cite{azerbayev2023proofnet} provides a key dataset for formalizing undergraduate-level mathematics. Subsequent efforts have diversified this landscape, including specialized benchmarks for combinatorics \cite{xiong2025combinatorial, liu2025combibench}, and the abstract algebra benchmark FATE \cite{jiang2025fate}.

\textbf{Auto-formalization with LLMs. } Early work by \cite{polu2020generative} pioneered the use of search-based methods for proof generation. Subsequent approaches have explored various search strategies, including variants of best-first search \cite{yang2023leandojo, lin2024lean, wu2025internlm2, li2024hunyuanprover,xin2025bfs} and Monte Carlo tree search \cite{lample2022hypertree, gloeckle2024abel, xin2024deepseek}. More recently, the field has seen a shift toward single-pass generation, as exemplified by state-of-the-art provers \cite{ren2025deepseek, lin2025goedel, wang2025kimina, zhou2025solving, chen2025seed}. These models are typically trained with large-scale reinforcement learning and generate long chain-of-thought reasoning to decompose and correct Lean code, achieving accuracy rates approaching 100\% on the miniF2F benchmark.

\textbf{Formalization of Different Types of Content. } The belief that all mathematical knowledge is amenable to formalization has motivated extensive efforts to formalize diverse domains. Researchers have developed frameworks and benchmarks for various types of content, though formalization efforts in theorem proving remain relatively less explored. For example, \cite{miranda2025veribench} formalized Python algorithms in VeriBench, \cite{liu2025beyond} introduced a framework for formalizing problem-solving processes, and \cite{bentkamp2023verified} contributed insights into the formalization of optimization problems.

\section{Formalization for Applied Mathematics}
\label{formalization_am}

While formal verification environments ensure rigorous correctness, standard benchmarks predominantly incentivize logical validity (proving a statement is true) over constructive synthesis (finding a solution). This creates a misalignment in evaluation tasks, which is common in applied mathematics. The primary objective of these problems is not to verify the existence of a solution, but to derive concrete symbolic or numerical results. To address this, we propose a versatile framework that flexibly adapts to computation-oriented tasks by leveraging Lean's rich constructive features, including definitions, inductive types, and type class instances. This approach enables the explicit modeling of solution structures and algorithms, ensuring that agents are evaluated on their ability to derive values for unknowns while maintaining compatibility with standard formal verification logic.

\subsection{Evaluation Problems}
\label{subsec:evaluation}

In applied mathematics, a prevalent class of problems involves explicit solution synthesis, such as solving linear equations, optimizing an objective function, or determining a probability distribution, rather than merely verifying the existence of such solutions. This distinction is fundamental: deriving a solution requires a constructive search process and numerical reasoning that extends beyond the scope of checking a candidate solution or proving existence non-constructively. Consequently, these evaluation problems represent a distinct paradigm from traditional theorem proving, necessitating a tailored formalization framework.

Consider the following quadratic optimization problem as a representative example:
\begin{align}
\label{optim_problem}
    \min_{x\in\mathbb{R}^n}\quad &f(x) = \frac 1 2x^T Ax - b^T x,
\end{align}
where $A \in \mathbb{R}^{n\times n}$ is a symmetric positive-definite matrix, and $b \in \mathbb{R}^n$ is a constant vector. The solution to the optimization problem is $x^*=A^{-1}b$. 
Previous approaches typically model such evaluation problems as existential propositions (i.e., proving there exists $x$, such that $x$ is an optimal point of problem \eqref{optim_problem})~\cite{liu2025beyond}. However, logical existence is a significantly weaker condition than constructive derivation. While identifying the explicit solution $x^*=A^{-1}b$ automatically entails its existence, establishing existence via non-constructive means yields no computational insight. For instance, the existence of an optimum can be proven via the extreme value theorem relying solely on domain compactness and function continuity without ever calculating the specific value of $x^*$. We observe that LLMs frequently exploit such abstract tactics to prove existence without constructing the explicit solution, as demonstrated by the case studies in Appendix~\ref{subsec:different_catrgories}. This misalignment allows LLMs to be credited with solving an evaluation problem absent any constructive solution, thereby undermining the benchmark's goal of assessing applied computational capabilities. To ensure robustness, we require a stricter representation that enforces explicit solution construction.

In the example above, the goal is to derive a constructive target variable $x^*$ expressed in terms of the given parameters $A$ and $b$, such that $x^*$ satisfies the optimality condition. Generalizing from this intuition, we propose the following formal definition.
\begin{definition}[Evaluation Problems]
    Given a property $Q(x, V)$ with parameters $V$, an evaluation problem is to construct an expression $x(V)$, such that $Q(x(V), V)$ holds.
\end{definition}

We implement this via a two-stage formalization pattern: 

\textbf{1. Construction. } An explicit function definition (def) is required to compute the solution term $x(V)$.

\textbf{2. Verification. } A subsequent theorem is required to prove that this specific term satisfies the property $Q(x(V), V)$. 
In the optimization example (\ref{optim_problem}), $V = \{A, b\}$, the constructive term is $x(V) = A^{-1}b$, and the property $Q(x, V)$ is the proposition that $x$ is an optimal point. The general formalization pattern is structured as follows.

\begin{tcolorbox}[
  colback=gray!10,
  colframe=black,
  boxrule=0.5pt,
  arc=1mm,
  left=2mm, right=2mm,
  top=0.2mm, bottom=0.2mm
]
\begin{lstlisting}[language=Lean]
variable {V : Type_v} {Q : Type_x → Type_v → Prop} {hV : Prop} 
        -- hV: assumptions for V

def x (V : Type_v) : Type_x := sorry

theorem verification (V : Type_v) :
        Q (x V) V := by sorry

\end{lstlisting}
\end{tcolorbox}

Crucially, this architecture enforces a strict syntactic barrier against non-constructive shortcuts. Although the model can read the full problem context, the definition of $x$ is rigidly constrained by its function signature: it accepts only the raw parameters $V$ as inputs, not the target property $Q$ or its proof. This structurally prevents the model from using the fact that a solution exists to construct the solution itself. Consequently, the agent is forced to perform explicit construction, building the solution term step-by-step from the available parameters.
See more examples in Appendix~\ref{subsec:different_catrgories}. 

\subsection{Algorithm Design Problems}

A significant portion of applied mathematics is concerned with defining algorithms. This encompasses iterative algorithms such as the gradient descent method for optimization, the Newton's method for optimization or root finding, or the alternating direction method of multipliers (ADMM) for large-scale learning. In these problems, the solution is not a closed-form expression, but a transition rule that evolves a state over discrete time steps to achieve a specific behavior (e.g., convergence to a stationary point). The objective here is to evaluate the agent's ability to synthesize the algorithmic logic itself, rather than merely verifying the properties of an abstract sequence.

Consider the classic gradient descent algorithm for minimizing a differentiable function $f(x)$. This algorithm is described mathematically by specifying an initialization $x_0$ and an iterative update rule:
\begin{align}
    x_{k+1} = x_k - \alpha_k \nabla f(x_k),
\end{align}
where $\alpha$ is the step size. The focus is on the constructive definition of how $x_{k+1}$ is derived from $x_k$.

To benchmark algorithm design effectively, we enforce a structural constraint: the agent must explicitly provide the loop body. We view the algorithm as a recursive function defined on natural numbers. The agent's task is to write the code for the $(n+1)$-th step given the $n$-th step. To formulate the mathematical intuition, we propose the following definition of an algorithm.

\begin{definition}[Algorithm]
    Given a state space $S$ and parameters $V$, constructing an algorithm requires defining a sequence generator $A: \mathbb{N} \to S$ such that:\\
    $A(0) = \text{init}(V)$ (Base Case),\\
    $A(n+1) = \Phi(A(n), V)$ (Inductive Step),\\
    where $\Phi$ is the transition function to be constructed.
\end{definition}

We formalize the algorithm design problems by modeling the structure of the iteration using Lean's recursive definition capabilities (matching on the inductive type \texttt{Nat}). This involves a two-stage process.

\textbf{Design.} 
The agent must define a function that maps the iteration number and current state to the next state. We formalize this as finding a transition function $\Phi: S \times V \to S$, where $S$ is the state space and $V$ are problem parameters.

\textbf{Analysis.} 
Once the sequence $A$ is effectively defined, the agent must prove a property $P(A)$, such as monotonicity ($f(A(n+1)) \le f(A(n))$) or convergence.

In our benchmark, this is implemented via the template below. The agent is strictly required to fill the first \texttt{sorry} with computable logic (the update rule) and the second \texttt{sorry} with the proof of correctness. The abstract formalization template is presented as follows.

\begin{tcolorbox}[
  colback=gray!10,
  colframe=black,
  boxrule=0.5pt,
  arc=1mm,
  left=2mm, right=2mm,
  top=0.2mm, bottom=0.2mm
]
\begin{lstlisting}[language=Lean]
variable {State : Type} (V : Type_v)
variable (P : (Nat → State) → Prop)

def algorithm (x0 : State) (V : Type_v) : Nat -> State
      | 0     := x0
      | (n+1) := sorry

theorem verification
    (x0 : State) (V : Type_v) :
    P (algorithm x0 V) := by sorry
\end{lstlisting}
\end{tcolorbox}

This framework shifts the evaluation from finding a number to writing a loop body, ensuring the agent acts as a true algorithm designer. See Appendix~\ref{subsec:different_catrgories} for additional examples.

\subsection{Representation Transformation}
\label{subsec:rep_trans}
In real-world applications, mathematical problems rarely present themselves in canonical forms. A critical capability of an applied mathematician is modeling: the ability to transform a raw, non-standard problem into a structured representation, such as integer linear programming, semidefinite programming, or a specific matrix factorization, that guarantees computational tractability. This process involves neither simple value derivation nor iterative algorithm design, but rather a structural transformation of the problem statement itself.

To evaluate this capability, we define a class of representation transformation problems. Here, the objective is to establish a rigorous mapping between a source problem instance and a target canonical form. We define a mathematical problem generically as a tuple $\mathcal{P} = (X, C, f)$, consisting of a decision variable $x$ in domain $X$, a set of constraints $C(x)$ (where $C: X \to \text{Prop}$), and an objective function $f: X \to V$, where $V$ represents the value space, such as $\mathbb{R}$ for optimization or a set of complex numbers $\mathbb{C}^n$ for eigenvalue problems. The goal of this task is to transform a source problem $\mathcal{P}_{src} = (X, C, f)$ into a target problem $\mathcal{P}_{tgt} = (Y, C', f')$, where $\mathcal{P}_{tgt}$ often belongs to a specific class of tractable problems such as convex problems. We formalize the relationship between these problems via the concept of relaxation.

\begin{definition}[Problem Relaxation and Equivalence]
We define a binary relation $\preceq$ on the space of optimization problems to formalize the structural hierarchy of reducibility. We say that $\mathcal{P}_{tgt}$ is a relaxation of $\mathcal{P}_{src}$, denoted $\mathcal{P}_{src} \preceq \mathcal{P}_{tgt}$, if every feasible solution in the source problem maps to a feasible solution in the target problem with an identical objective value:
\begin{equation}
\forall x \in X, C(x) \implies \exists y \in Y, C'(y) \land f(x) = f'(y).
\end{equation}
Two problems are equivalent if they are mutually comparable under this relation, satisfying the antisymmetric property: $\mathcal{P}_{src} \preceq \mathcal{P}_{tgt} \land \mathcal{P}_{tgt} \preceq \mathcal{P}_{src}$.
\end{definition}

This formalization captures the essence of mathematical modeling. For example, proving that a specific graph cut problem is equivalent to a spectral clustering formulation requires showing that the set of cuts maps to the spectrum of the Laplacian matrix (where $f$ maps the graph configuration to eigenvalues). In Lean, we operationalize this by requiring the agent to defining the \texttt{target} problem structure and proving the relaxation or equivalence relation. The framework distinguishes between the source problem (given as parameters) and the target form (which the agent must construct).

\textbf{Modeling (Construction).} The agent must define the components of the target problem: the new domain $Y$, constraints $C'$, and the objective $f'$.

\textbf{Verification.} The agent must prove the logical proposition $\mathcal{P}_{src} \preceq \mathcal{P}_{tgt}$ (or equivalence).

To rigorize this process, we first define a generic MathProblem structure that encapsulates the decision domain $D$, the constraint predicate, and the objective function. Central to this framework is the relaxes definition, which formalizes the notion of problem reduction: it asserts that for every feasible solution in the source problem $P_1$, there exists a mapped solution in the target problem $P_2$ yielding the same objective value. In this task, the agent is presented with a source problem P\_src and must fulfill three specific obligations to resolve the \texttt{sorry} placeholders: construct the target domain D\_tgt (often requiring the introduction of auxiliary slack variables), explicitly define the target problem P\_tgt, and prove the valid\_transformation theorem, which demands a demonstration of mutual relaxation between the source and target problems. The generalized schema for representation transformation is shown below. The agent acts as a translator, converting a user-defined problem into a canonical form required by a solver. 

\begin{tcolorbox}[
colback=gray!10,
colframe=black,
boxrule=0.5pt,
arc=1mm,
left=2mm, right=2mm,
top=0.2mm, bottom=0.2mm
]
\begin{lstlisting}[language=Lean]
structure MathProblem (D V : Type) :=
    (constraints : D → Prop)
    (objective : D → V)

def relaxes {D1 D2 V : Type}
    (P1 : MathProblem D1 V)
    (P2 : MathProblem D2 V) : Prop :=
    ∀ x, P1.constraints x →
    ∃ y, P2.constraints y ∧ P1.objective x = P2.objective y
    
variable {D_src V : Type}
    (P_src : MathProblem D_src V)
def D_tgt : Type := sorry
def P_tgt : MathProblem D_tgt V := sorry

theorem valid_transformation :
    relaxes P_src P_tgt ∧ relaxes P_tgt P_src := by sorry
\end{lstlisting}
\end{tcolorbox}

By structuring benchmarks this way, we ensure that the evaluation metric aligns with the mathematical intuition of problem reduction. The agent is not merely asked to solve the problem, but to demonstrate that it understands the mathematical structures well enough to translate them across different domains while preserving their fundamental properties. See Appendix~\ref{subsec:different_catrgories} for case studies.

\subsection{Theorem Proving}
While our benchmark emphasizes constructive tasks, theorem proving remains a cornerstone of formal verification. We include a subset of traditional theorem proving tasks to serve as a baseline for assessing the logical reasoning capabilities of models distinct from their constructive abilities. Unlike the previous categories where the solution is a data structure or algorithm, here the objective is purely deductive. Given a proposition $P$ in a context $\Gamma$, the task is to construct a proof term $t$ such that $\Gamma \vdash t : P$. In Lean 4, this follows the standard declaration pattern.

\begin{tcolorbox}[
colback=gray!10,
colframe=black,
boxrule=0.5pt,
arc=1mm,
left=2mm, right=2mm,
top=0.2mm, bottom=0.2mm
]
\begin{lstlisting}[language=Lean]
theorem proposition_name (V : Parameters) (h : Assumptions) :
  Target_Proposition := by sorry
\end{lstlisting}
\end{tcolorbox}

The agent must replace \texttt{sorry} with a valid tactic sequence or proof term that closes the goal. These problems range from establishing properties of convex sets to proving inequalities in high-dimensional probability, serving as a control group to measure if models trade off pure deductive power for constructive flexibility.

\section{Benchmark Curation and Characteristics}
\label{sec:benchmark_curation}

\subsection{Benchmark Curation}

\paragraph{Problem Collection.} Content selection prioritizes the benchmark’s representativeness and domain-specific characteristics, with samples predominantly sourced from classical and well-recognized textbooks across convex analysis, optimization, numerical linear algebra, and high-dimensional probability~\cite{boyd2004convex, golub2013matrix, vershynin2018high}.
The selection is conducted by doctoral and senior undergraduate students specializing in applied mathematics, who identify topics representative of core concepts.
A preliminary difficulty classification based on problem sources serves as a baseline: easy problems correspond to standard course exercises, while hard problems align with PhD qualifying examinations.
We categorize samples into theorem proving and construction-verification. The latter requires synthesizing a solution before verification across evaluation, algorithm design, and representation transformation tasks. Detailed definitions appear in Section~\ref{formalization_am}. Comprehensive statistics regarding the task distribution across domains are provided in Appendix~\ref{sec:benchmark_analysis}. By incorporating both task types, our benchmark is designed to evaluate a model's proficiency in varying mathematical reasoning modalities—ranging from the rigorous derivation required in probability theory to the algorithmic synthesis essential in optimization and numerical algebra.

\paragraph{Lean Formalization.} 
Formalization is executed by a team of experienced Lean users. Leveraging a standardized template to ensure strict adherence to Mathlib norms, the team converts 50 selected problems from each subfield into executable Lean code. This rigorous process guarantees that the resultant code maintains strong logical consistency with the original mathematical statements and exhibits excellent reproducibility in standard Lean 4 environments.

\paragraph{Review.}
\label{subsec:review}
Validation employs a rigorous two-stage process to ensure strict semantic consistency between the formalized Lean code and the original mathematical problems, consisting of an LLM-based preliminary check and manual in-depth verification.
In the LLM preliminary check, we use a state-of-the-art LLM to automatically screen the entire corpus, checking each formalized problem against its original statement. The LLM compares the formalized code with the original problem statements to detect possible semantic discrepancies, logical omissions, or misalignments; all flagged cases are prioritized for manual review and revision. 
Detailed information regarding the specific model, configuration, and prompting strategies is provided in the Appendix~\ref{LLM_Verification_Details}.
The manual in-depth verification adopts several rounds of team cross-validation, with each expert being an experienced Mathlib contributor. In each round, every formalized problem is reviewed by at least two Lean experts, who strictly verify that all constructs, both custom definitions/lemmas and invoked Mathlib concepts, align with the original problem’s intent.

\subsection{Benchmark Characteristics}
\paragraph{Expert-Driven Difficulty Assessment.}
Unlike traditional benchmarks that may rely on syntactic length or heuristic proxies for difficulty, we ground our classification in expert mathematical consensus. To measure the intrinsic logical gap between hypothesis and conclusion, we conduct an assessment with ten PhD students from prestigious research institutes. They evaluate the reasoning depth required to construct a proof blueprint, ensuring that problems classified as ``hard'' reflect genuine mathematical complexity (e.g., requiring novel auxiliary constructions) rather than mere notational verbosity. This human-centric approach validates that our AMBER benchmark captures the rigor of research-level applied mathematics.

\begin{figure}[ht]
\centering
\centerline{\includegraphics[width=0.95\columnwidth]{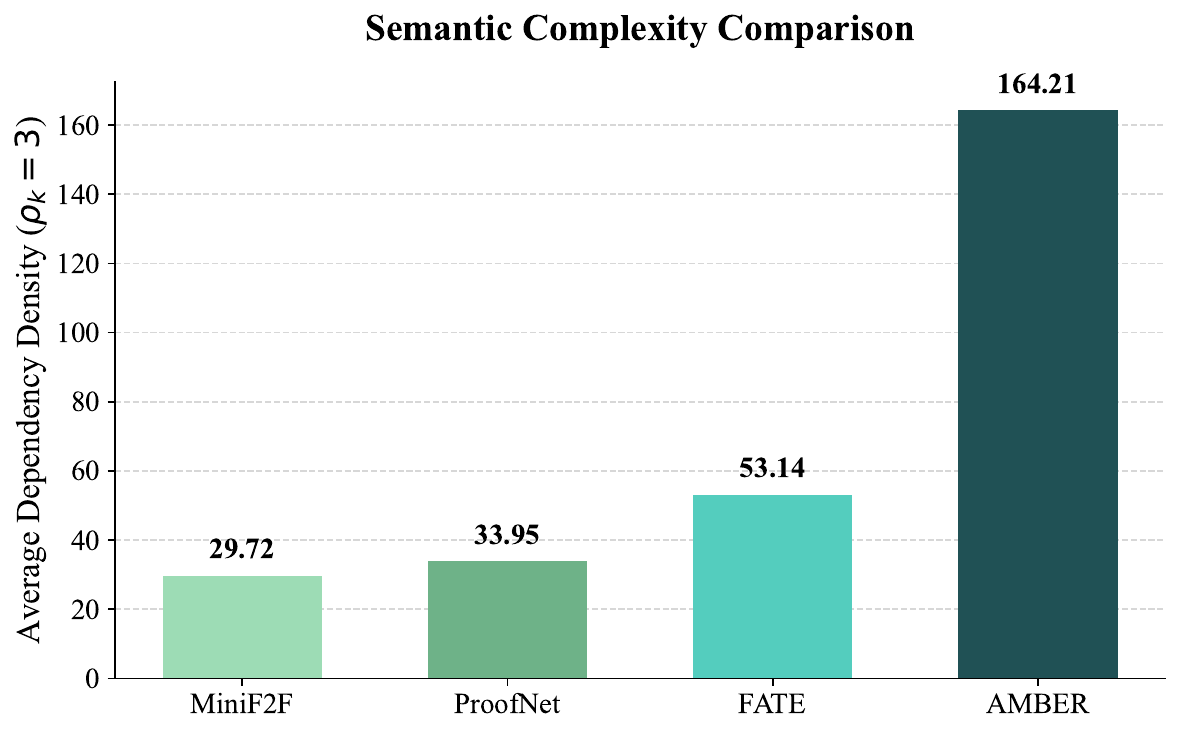}}
\caption{Quantitative comparison of semantic complexity.}
\label{fig:complexity_comparison}
\end{figure}

\paragraph{Quantitative Formal Complexity.}
Complementing qualitative expert assessment, we quantify the intrinsic information density of our benchmark using Lean's metaprogramming framework. Unlike text-based datasets where complexity is often proxied by token length, formal environments allow us to measure the conceptual dependency load of a problem statement.
Concretely, we represent each benchmark instance $p$ as the corresponding Lean declaration and model the underlying formal library as a global semantic dependency graph $\mathcal{G}$ with vertex set $\mathcal{V}$ and directed edge set $\mathcal{E}$, where $\mathcal{V}$ consists of formal declarations such as definitions and theorems, and $\mathcal{E}$ encodes usage dependencies. For a problem $p$, we define the dependency density $\rho_k(p)$ as the size of its non-trivial conceptual neighborhood:
\[
\rho_k(p) = \Big| \{ v \in \mathcal{V} \mid 1 \le \text{dist}_{\mathcal{G}}(p, v) \le k \land v \notin \mathcal{S}_{\text{trivial}} \} \Big|,
\]
where $\text{dist}_{\mathcal{G}}(p, v)$ denotes the shortest path distance from problem $p$ to concept $v$ in the dependency graph.
The set $\mathcal{S}_{\text{trivial}}$ acts as a high-pass semantic filter, removing purely syntactic constructs such as logic connectives and basic types to isolate domain-specific knowledge.
We set the local radius $k=3$ to capture the immediate conceptual prerequisites required to comprehend the statement, avoiding the dilution of metrics by foundational axioms found at deeper levels. A higher $\rho_k(p)$ indicates that the model must successfully retrieve and reason over a dense subgraph of mathematical concepts to even parse the problem correctly, posing a significant challenge to the model's retrieval-augmented generation (RAG) and long-context understanding capabilities. To quantify the overall complexity of a dataset $\mathcal{D}$, we calculate the average dependency density as the arithmetic mean of the individual densities: $\bar{\rho}_k(\mathcal{D}) = \frac{1}{|\mathcal{D}|} \sum_{p \in \mathcal{D}} \rho_k(p)$.
To contextualize these metrics, we compare AMBER against three established baselines: MiniF2F (high-school competitions), ProofNet (undergraduate mathematics), and FATE. 
Figure~\ref{fig:complexity_comparison} presents the results. Our AMBER benchmark exhibits the highest dependency density, quantitatively confirming the superior semantic depth of AMBER.

\section{Experiments}
\begin{table*}[t]
\centering
\caption{Pass@16 results of various models across different subsets. Each dataset is partitioned into Easy and Hard subsets. }
\label{tab:results}
\begin{tabular}{l cc cc cc cc}
\toprule
& \multicolumn{2}{c}{\textbf{Optimization}} & \multicolumn{2}{c}{\textbf{Convex Analysis}} & \multicolumn{2}{c}{\textbf{Numerical Algebra}} & \multicolumn{2}{c}{\textbf{Probability}} \\
\cmidrule(lr){2-3} \cmidrule(lr){4-5} \cmidrule(lr){6-7} \cmidrule(lr){8-9}
\textbf{Model} & Easy & Hard & Easy & Hard & Easy & Hard & Easy & Hard \\
\midrule
DeepSeek-V3.2-Thinking   & 3/24 & 0/26 & 5/26 & 0/24 & 3/25 & 0/25 & 1/25 & 0/25 \\
GPT-5.1       & 2/24 & 0/26 & 3/26 & 0/24 & 3/25 & 0/25 & 1/25 & 0/25 \\
Gemini-3.0 Pro & 3/24 & 0/26 & 4/26 & 0/24 & 4/25 & 0/25 & 1/25 & 0/25 \\
\midrule
Goedel Prover-32B & 3/24 & 0/26 & 1/26 & 0/24 & 3/25 & 0/25 & 1/25 & 0/25 \\
Kimina Prover-72B & 2/24 & 0/26 & 2/26 & 0/24 & 3/25 & 0/25 & 1/25 & 0/25 \\

\bottomrule
\end{tabular}
\end{table*}

To assess the capabilities of current models on applied mathematics formalization, we conduct a comprehensive evaluation on our AMBER benchmark. Our experiments focus not only on theorem proving, but also on agents' ability to follow the structural requirements, specifically completing various forms of Lean content including evaluation, algorithm design and representation transformation, outlined in Section \ref{formalization_am}.

\subsection{Experiment Setup}
\label{subsec:experiment_setup}
\paragraph{Models.} We evaluate a diverse set of models representing two distinct paradigms: general-purpose reasoning agents and specialized theorem provers.
For general reasoning, we employ DeepSeek-V3.2-Thinking \cite{liu2025deepseek} and GPT-5.1, which represent state-of-the-art capabilities in natural language understanding and chain-of-thought reasoning.
For specialized formalization, we evaluate Goedel Prover-32B~\cite{lin2025goedel} and Kimina Prover-72B~\cite{wang2025kimina}, models that have been fine-tuned extensively on Mathlib and Lean corpora.

\paragraph{Evaluation Protocol.} 
For each problem, the model is provided with a context header (imports and variable declarations) and a Lean statement based on the structural template corresponding to the problem type, including evaluation, algorithm design, representation transformation and traditional theorem proving. Crucially, the model must complete the code by filling in the \texttt{sorry} placeholders for both the definition (construction) and the theorem (verification).
We report the Pass@$k$ metric ($k=16$). A problem is considered solved only if, within $k$ samples, at least one generated solution allows Lean to successfully compile both the constructive definition and the verification proof without errors. Unbiased estimators for Pass@$k$ with varying $k$ are detailed in Appendix~\ref{subsec:unbiased}. Additional hyperparameters and prompt engineering details are provided in Appendix~\ref{sec:experiment_details}.

\subsection{Benchmark Performance}
\label{subsec:performance}
We summarize the performance across four applied mathematics subfields in Table~\ref{tab:results}. The results reveal several critical insights regarding the current state of automated formalization, most notably that overall pass rates are remarkably low compared to standard pure mathematics benchmarks like MiniF2F. This performance gap is particularly evident in the hard subset, where success rates remain near zero for almost all evaluated models. These findings validate our hypothesis that applied mathematics presents unique structural challenges. Agents must overcome both the intrinsic logical depth required to navigate specialized domain-specific proofs and the significant structural complexity involved in constructing accurate mathematical objects. This difficulty is consistent with our quantitative analysis, which shows our AMBER benchmark possesses a significantly higher semantic dependency density than existing datasets, requiring models to reason over a much denser subgraph of mathematical concepts. We provide a comprehensive breakdown of unbiased Pass@$k$ results for $k \leq 16$ in Table~\ref{tab:pass_k_simp}, with full results detailed in Appendix~\ref{subsec:unbiased}.

\begin{table}[h]
\centering
\caption{Pass@k ($k=1,16$) statistics for all models on the full AMBER benchmark. The values represent the percentage of problems solved given $k$ budget.}
\label{tab:pass_k_simp}
\begin{tabular}{lcc}
\toprule
\textbf{Model} & \textbf{Pass@1} & \textbf{Pass@16} \\
\midrule
\multicolumn{3}{l}{\textit{General-Purpose Reasoning Models}} \\
DeepSeek-V3.2-Thinking & 1.68\% & 6.00\% \\
GPT-5.1 & 1.40\% & 4.50\% \\
Gemini-3.0 Pro & 2.62\% & 6.00\% \\
\midrule
\multicolumn{3}{l}{\textit{Specialized Theorem Provers}} \\
Goedel Prover-32B & 0.97\% & 4.00\% \\
Kimina Prover-72B & 1.03\% & 4.00\% \\
\bottomrule
\end{tabular}
\end{table}

\textbf{General Models Outperform Specialized Provers.}  A striking observation is that general-purpose reasoning models consistently outperform specialized mathematical provers, a trend that contradicts the typical performance hierarchy where specialized provers dominate standard benchmarks. We attribute this to a phenomenon we term ``tactical overfitting": while extensive fine-tuning on state-tactic pairs maximizes the ability to close logical goals, it appears to induce a ``hammer-and-nail" bias where models treat every structural challenge as a pure theorem-proving task. This specialization leads to a catastrophic forgetting of general alignment skills, causing models to ignore the multi-stage nature of applied mathematics. Consequently, they struggle to adhere to the rigid ``construction-verification'' template, frequently bypassing the required definition (the \texttt{def} block) or erroneously attempting to discharge construction goals using non-constructive proof tactics. In contrast, general-purpose models retain robust instruction-following skills, allowing them to bridge the gap between informal intent and formal syntax to correctly synthesize mathematical objects before entering the verification stage. A Detailed analysis of this phenomenon is provided in Appendix~\ref{sec:details_instruction}.

\subsection{Ablation Study on Prompts.} 
\label{subsec:ablation_prompt}
To validate the effectiveness of our proposed framework and the necessity of structural constraints, we conducted an extensive ablation study across all 100 problems within the optimization and convex analysis domains. We compared our construction-verification prompt against two baselines: standard code completion (providing the file without structural instructions) and traditional proof only (asking solely for a proof). The results, detailed in Appendix~\ref{subsec:ablation}, demonstrate that the construction-verification strategy yields the highest success rate of 8\%. In contrast, the performance drops to 6\% with standard completion and further to 5\% with proof-only prompts.

\begin{table}[htbp]
\centering
\caption{Ablation results on the 100-problem subset.}
\label{tab:ablation_prompts}
\begin{tabular}{lcc}
\toprule
\textbf{Prompt Type} & \textbf{Success Rate}  \\
\midrule
Construction-Verification & \textbf{8/100} \\
Standard Completion & 6/100   \\
Traditional Proof-Only & 5/100  \\
\bottomrule
\end{tabular}
\end{table}

Qualitative analysis reveals that traditional proof-only prompts frequently mislead models into using non-constructive shortcuts. For instance, models often resort to tactics such as \texttt{Classical.choice} to prove existence logically without providing the computational solution. Furthermore, standard completion prompts often lead models to ignore the requirement to fill the definition block and attempt to discharge the entire problem using only proof-search tactics. These findings confirm that explicit structural prompting is essential to prevent models from bypassing the constructive requirements of the benchmark.

\subsection{Error Analysis}
\label{subsec:error_analysis}
To investigate the specific obstacles in formalizing applied mathematics, we conducted a detailed manual classification of 3,189 randomly sampled error messages generated by the DeepSeek-R1 model. The analysis reveals that the challenges are not uniformly distributed, with specific failure modes dominating the results.

\textbf{Hallucination of LLMs.} The most prevalent source of error, accounting for 47.3\% (1,512 cases) of the total, is the invocation of non-existent theorems or definitions. In these instances, the model hallucinates plausible-sounding Mathlib identifiers (e.g., \texttt{convexHull\_eq\_sInter}) that do not exist in the actual library. This indicates that while the model often correctly identifies the necessary mathematical steps, it lacks the precise memorization of the Lean 4 library signature required to implement them.

\textbf{Formalization Competence.} The second largest category, comprising the remaining 32.7\% of errors, stems from insufficient formalization skills. These failures occur when the model utilizes valid theorems and tactics but misapplies them, misjudging the effect of a tactic or failing to discharge necessary side-conditions for a rewrite. This reflects a gap between the model's high-level reasoning and its low-level tactic execution capabilities.

\textbf{Incomplete Outputs.} Despite explicit instructions to provide fully executable code, 15.2\% (487 cases) of the outputs contained incomplete segments marked by \texttt{sorry} or \texttt{submit}. This issue is particularly pronounced in complex algorithm design tasks, where the model successfully sketches the algorithmic structure but abandons the rigorous proof of the inductive step, reverting to placeholders when the derivation becomes too involved.

\textbf{Formatting and Syntax.} Finally, a minor fraction of 4.8\% (154 cases) is attributed to basic output formatting errors, such as failing to encapsulate code within standard markdown blocks or generating syntactically malformed headers that prevent the compiler from parsing the file.

In summary, nearly half of the failures are due to LLM hallucination, suggesting that future improvements should prioritize retrieval-augmented generation or specialized fine-tuning to align the model's internal knowledge with the exact state of the Mathlib library. Details and case studies are provided in Appendix~\ref{sec:details_error}.

\begin{figure}[ht]
    \centering
    \includegraphics[width=0.8\columnwidth]{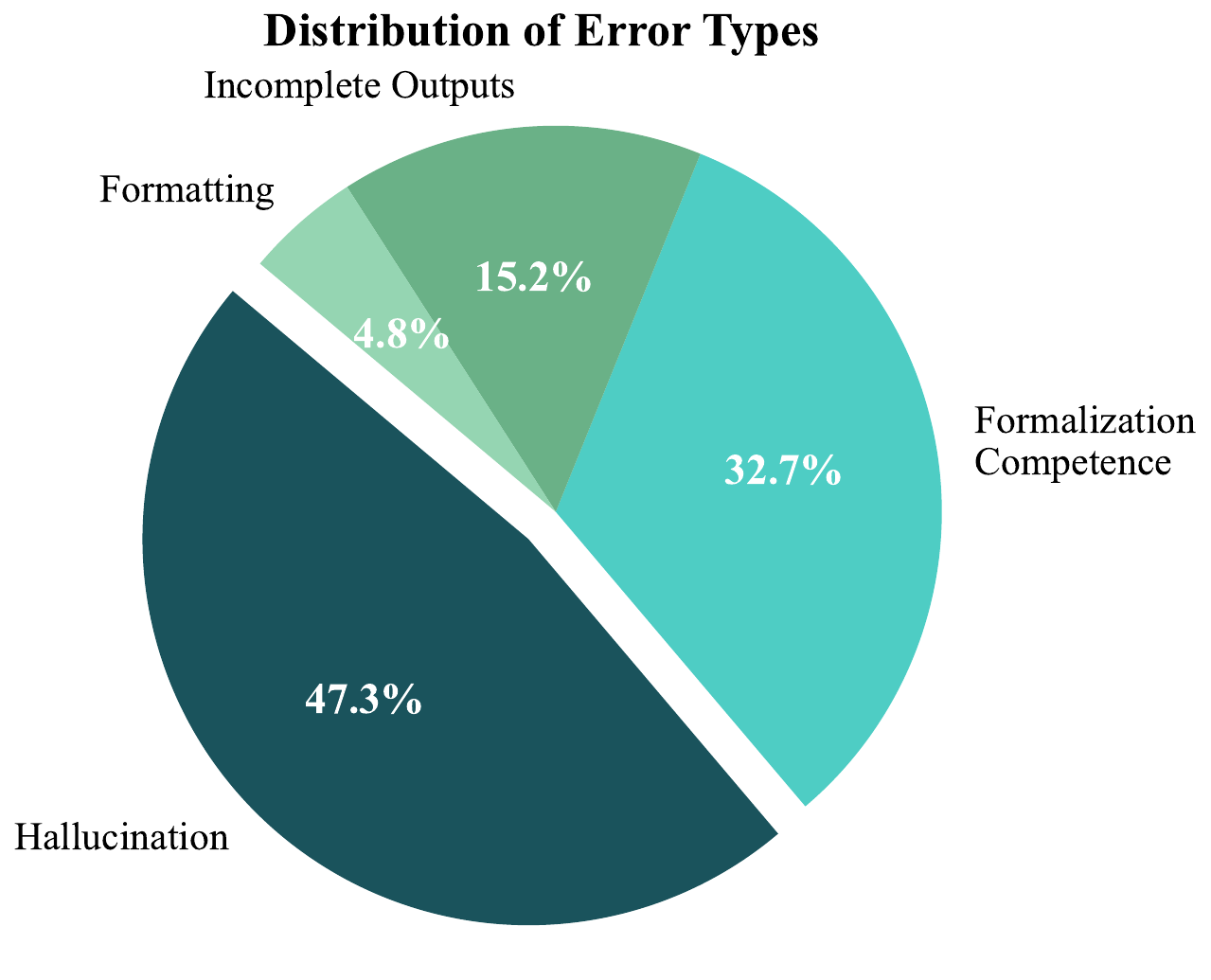}
    \caption{Error analysis distribution.}
    \label{fig:error_analysis}
\end{figure}

\section{Conclusion}
\label{sec:conclusion}

In this work, we introduce a comprehensive AMBER benchmark tailored for applied mathematics in Lean 4, addressing a critical misalignment in existing datasets by shifting the evaluation paradigm from existential verification to explicit solution construction. Apart from theorem proving, our proposed tasks spanning numerical evaluation, algorithm design and representation transformation compel agents to go beyond mere theorem proving by synthesizing computable definitions and performing structural modeling before verification. Extensive evaluations reveal that this constructive paradigm poses significant challenges to state-of-the-art models, most notably highlighting a ``tactical overfitting'' phenomenon where specialized theorem provers, despite their proficiency in closing local logical goals, exhibit a rigidity that hampers their ability to adhere to multi-stage constructive instructions compared to the superior versatility of general-purpose models. These results underscore the necessity of moving beyond pure proof search towards neuro-symbolic systems that integrate high-level algorithmic reasoning with rigorous formal verification, serving as a foundation for future research in expanding the frontiers of formalized applied mathematics.

\bibliography{ref}
\bibliographystyle{icml2026}

\newpage

\appendix
\onecolumn

\section{Introduction to Formalization and Lean}
\label{sec:intro_to_lean}

Mathematical formalization is the process of translating mathematical definitions, theorems, and proofs from natural language (informal mathematics) into a strict formal language that can be mechanically checked by a computer. This appendix provides a brief overview of the Lean theorem prover, focusing on the concepts essential for understanding the benchmark presented in this paper.

\subsection{Formal Verification and Lean}
Lean is an interactive theorem prover (ITP) and a functional programming language developed at Microsoft Research. Unlike computer algebra systems (e.g., MATLAB, Mathematica) which focus on numerical or symbolic computation, Lean focuses on logical correctness. It verifies that a derivation follows the axioms of mathematics without ambiguity.

The core underlying logic of Lean is the calculus of inductive constructions (CIC). In this system, mathematical statements and proofs are treated as types and programs, respectively. This correspondence, known as the Curry-Howard Isomorphism, implies that:
\begin{itemize}
\item A proposition (a mathematical statement) is a type.
\item A proof of that proposition is a term (a program) of that type.
\item Verifying a proof is equivalent to type-checking the program.
\end{itemize}

If a user writes a proof term that successfully type-checks against a theorem statement, the theorem is guaranteed to be true (relative to the consistency of the system).

\subsection{Definitions vs. Theorems}
\label{subsec:def_vs_thm}
A critical distinction in Lean and a central theme of our AMBER benchmark is the difference between defining data (definitions) and proving facts (theorems). In applied mathematics, we often need both: we construct a solution (an algorithm or a value) and then prove it is correct.

\textbf{Definitions (\texttt{def}).} Definitions are used to construct mathematical objects, functions, or algorithms. These are computational or structural entities. For example, defining the square function on natural numbers:
\begin{tcolorbox}[
  colback=gray!10,
  colframe=black,
  boxrule=0.5pt,
  arc=1mm,
  left=2mm, right=2mm,
  top=0.2mm, bottom=0.2mm
]
\begin{lstlisting}[language=Lean]
def square (n : ℕ) : ℕ := n * n
\end{lstlisting}
\end{tcolorbox}

In our benchmark, the construction phase (e.g., defining the optimal $x^*$, or writing the Gradient Descent loop) corresponds to the creation of a \texttt{def}.

\textbf{Theorems (\texttt{theorem}).}
Theorems are used to assert propositions. They do not produce data; they carry logical evidence. For example, proving that the square of a positive number is positive:
\begin{tcolorbox}[
  colback=gray!10,
  colframe=black,
  boxrule=0.5pt,
  arc=1mm,
  left=2mm, right=2mm,
  top=0.2mm, bottom=0.2mm
]
\begin{lstlisting}[language=Lean]
theorem square_pos (n : ℕ) (h : n > 0) : square n > 0 := by
    -- The proof goes here
    sorry
\end{lstlisting}
\end{tcolorbox}

In our benchmark, the verification phase corresponds to establishing a \texttt{theorem} that validates the properties of the previously defined \texttt{def}.

\subsection{Tactics and Automation}
Writing proof terms directly (e.g., constructing complex lambda calculus terms) is tedious for humans. Lean provides a tactic mode, usually introduced by the keyword \texttt{by}. Tactics are meta-programs that manipulate the proof state to generate the underlying proof term automatically.

Common tactics include:\\
\texttt{rw} (rewrite): Replaces terms using equalities.\\
\texttt{simp} (simplify): Simplifies expressions using a database of algebraic rules.\\
\texttt{apply}: Applies a lemma to reduce the current goal to sub-goals.\\
\texttt{linarith}: Automatically solves linear arithmetic inequalities.

In our experiments, LLMs must predict these tactic sequences to close the logical goals. The \texttt{sorry} keyword is a placeholder used to admit a goal without proof. The objective of the benchmark is to replace every \texttt{sorry} with valid code or tactics.

\subsection{Mathlib: The Mathematical Library}
Lean is supported by Mathlib, a unified, community-driven mathematical library. Mathlib contains formalized definitions and theorems spanning algebra, topology, analysis, probability, and more.
For applied mathematics, Mathlib provides the necessary infrastructure, such as:\\
\textbf{Linear Algebra:} Definitions of matrices, vectors, eigenvalues, and inner product spaces.\\
\textbf{Analysis:} Notions of derivatives, gradients, convexity, and continuity.\\
\textbf{Probability:} Measure theory, integration, and random variables.

Our AMBER benchmark heavily relies on Mathlib. For instance, an optimization problem in our dataset will invoke Mathlib's definition of PositiveDefinite matrices. An LLM's performance is thus contingent not only on its reasoning ability but also on its familiarity with Mathlib's specific API and naming conventions.

\subsection{Why Lean for Applied Mathematics?}
While historically used for pure mathematics (e.g., verifying Fermat's Last Theorem), Lean 4 is designed as a general-purpose programming language. This makes it uniquely suitable for the Algorithmic and Computational problems featured in this paper.

Unlike purely logical systems, Lean allows us to define executable algorithms, run these algorithms on inputs and even simultaneously prove that the algorithm converges.This duality allows our benchmark to evaluate agents on the full spectrum of applied mathematics: from modeling and computation (programming) to rigorous justification (proving).
\section{Details and Examples on Formalization Framework}
\label{sec:framework}
In this section, we provide details on the formalization framework in Section~\ref{formalization_am}. We first show the detailed \texttt{class} definition

\subsection{Classes of Representation Transformation}
\label{subsec:class_transformation}
To systematically evaluate the capability of models to perform Representation Transformation, we have developed a comprehensive infrastructure of modular Type Classes in Lean 4. This framework is designed to rigorously formalize the often implicit process of problem reformulation in applied mathematics, converting it from a heuristic procedure into a precise task of type mapping. Unlike standard numerical computation, representation transformation demands the restructuring of a non-standard primal problem into a computationally tractable canonical form, such as LP or SDP, while preserving mathematical equivalence or relaxation properties. To this end, we established a hierarchical class structure. We begin by defining a generic interface, OptProblem, which encapsulates the decision variable domain, the set of constraint predicates, and the objective function. Building upon this foundation, and specifically addressing duality theory, one of the cornerstones of optimization theory , we designed specialized class structures to formalize the derivation of dual problems. This design not only encapsulates the construction of the Lagrangian function but also explicitly types the space of dual variables and their associated constraints. By enforcing this structured definition, we compel the agents to go beyond intuitive symbolic derivation; they must instead construct a verified structural transformation, providing rigorous proofs of the correctness within the Lean type system.
\newpage
\begin{tcolorbox}[
  colback=gray!10,
  colframe=black,
  boxrule=0.5pt,
  arc=1mm,
  left=2mm, right=2mm,
  top=0.2mm, bottom=0.2mm
]
\begin{lstlisting}[language=Lean]
class OriginalProblem where
  n_var : ℕ
  constraints : (Fin n_var → ℝ) → Prop
  objective : (Fin n_var → ℝ) → ℝ

class OptProblem extends OriginalProblem where
  n_eqs : ℕ
  eq_constraints : (Fin n_var → ℝ) → (Fin n_eqs → ℝ)
  n_ieqs : ℕ
  ieq_constraints : (Fin n_var → ℝ) → (Fin n_ieqs → ℝ)
  constraints := fun x => eq_constraints x = 0 ∧ ieq_constraints x ≤ 0
  h_constraints : constraints =  fun x => eq_constraints x = 0 ∧ ieq_constraints x ≤ 0 := by simp

class LP extends OptProblem where
  c : Fin n_var → ℝ
  A_eq : Matrix (Fin n_eqs) (Fin n_var) ℝ
  b_eq : Fin n_eqs → ℝ
  A_ieq : Matrix (Fin n_ieqs) (Fin n_var) ℝ
  b_ieq : Fin n_ieqs → ℝ
  objective := fun x => c ⬝ᵥ x
  eq_constraints := fun x => A_eq *ᵥ x - b_eq
  ieq_constraints := fun x => A_ieq *ᵥ x - b_ieq
  h_objective : objective = fun x => c ⬝ᵥ x := by simp
  h_eq : eq_constraints = fun x => A_eq *ᵥ x - b_eq := by simp
  h_ieq : ieq_constraints = fun x => A_ieq *ᵥ x - b_ieq := by simp

class SDP extends OriginalProblem where
  c : Fin n_var → ℝ
  n_eqs : ℕ
  A_eq : Matrix (Fin n_eqs) (Fin n_var) ℝ
  b_eq : Fin n_eqs → ℝ
  eq_constraints := fun x => A_eq *ᵥ x - b_eq
  n_ieqs : ℕ
  A_sd : Fin n_var → Matrix (Fin n_ieqs) (Fin n_ieqs) ℝ
  B_sd : Matrix (Fin n_ieqs) (Fin n_ieqs) ℝ
  ieq_constraints := fun x => ∑ i, x i • A_sd i + B_sd
  constraints := fun x => eq_constraints x = 0 ∧ (ieq_constraints x).PosDef
  h_constraints : constraints =  fun x => eq_constraints x = 0 ∧ (ieq_constraints x).PosDef := by simp
  objective := fun x => c ⬝ᵥ x
  h_objective : objective = fun x => c ⬝ᵥ x := by simp

def subequivlance (p q : OriginalProblem) : Prop :=
  ∀ (x : Fin p.n_var → ℝ), (p.constraints x) →
  ∃ (y : Fin q.n_var → ℝ), (q.constraints y) ∧
  q.objective y = p.objective x

def equivalence (p q : OriginalProblem) : Prop :=
  subequivlance p q ∧ subequivlance q p

class DualProblem (p : OptProblem) where
  dual_objective : (Fin p.n_eqs → ℝ) → (Fin p.n_ieqs → ℝ) → EReal
  dual_domain : Set ((Fin p.n_eqs → ℝ) × (Fin p.n_ieqs → ℝ))
  h_objective : dual_objective = fun lam mu => (⨅ x : (Fin p.n_var → ℝ), ((lam ⬝ᵥ p.eq_constraints x) + (mu ⬝ᵥ p.ieq_constraints x) + p.objective x).toEReal) := by simp
  h_domain : dual_domain = {(lam, mu) | dual_objective lam mu ≠ ⊥} := by simp


\end{lstlisting}
\end{tcolorbox}

\subsection{Examples of Different Problem Categories}
\label{subsec:different_catrgories}
In this section, we provide concrete examples illustrating the distinct problem categories defined in our AMBER benchmark.
\subsubsection{Construction-Verification vs. Theorem-proving}

Construction-Verification problems ask the solver to produce an explicit value of a mathematics problem, rather than merely proving that such a value exists. Below is a representative example of ..........

\textbf{Informal Statement.} Solve the optimization problem
\[
f = \|x\| , \text{ s.t. } x \in \left\{(x,y,z) | x^2+y^2+z^2 \le 1\right\}.
\]

The theorem-proving approach is to translate optimization problem into an existence theorem , and then extract a witness using \texttt{use} tactic as follow.

\begin{tcolorbox}[
  colback=gray!10,
  colframe=black,
  boxrule=0.5pt,
  arc=1mm,
  left=2mm, right=2mm,
  top=0.2mm, bottom=0.2mm
]
\begin{lstlisting}[language=Lean]
noncomputable def f : (EuclideanSpace ℝ (Fin 3)) → ℝ :=
  fun x ↦ ∥x∥

def ball := Metric.closedBall (0 : EuclideanSpace ℝ (Fin 3)) 1

theorem optimization_problem' : ∃ x ∈ ball, IsMinOn f ball x := by
  use 0
  constructor
  · simp [ball]
  · intro x hx
    simp [f]
\end{lstlisting}
\end{tcolorbox}

In practice, Lean code generated by LLMs often default to an automation-friendly pattern: they discharge goals by rewriting them into standard existence statements (e.g., via extreme-value arguments), yielding a proof of $\exists x,\, \dots$ rather than an explicit construction of a concrete minimizer. Consequently, even when the underlying optimization problem admits a simple closed-form solution, the generated proof script may drift toward establishing the existence of some \texttt{x} instead of defining a specific \texttt{x} and certifying its feasibility and optimality. In the example below, the proof establishes only the existence of the minimizer \texttt{x} by showing that the feasible set is compact, without providing an explicit construction of the minimizer \texttt{x}.
\begin{tcolorbox}[
  colback=gray!10,
  colframe=black,
  boxrule=0.5pt,
  arc=1mm,
  left=2mm, right=2mm,
  top=0.2mm, bottom=0.2mm
]
\begin{lstlisting}[language=Lean]
noncomputable def f : (EuclideanSpace ℝ (Fin 3)) → ℝ :=
    fun x ↦ ∥x∥

def ball := Metric.closedBall (0 : EuclideanSpace ℝ (Fin 3)) 1

theorem optimization_problem {n : ℕ} [NeZero n] :
      ∃ x ∈ ball, IsMinOn f ball x :=
    -- Show that ball is a Compact set
    have hcompact : IsCompact ball :=
        isCompact_closedBall 0 1
    hcompact.exists_isMinOn (by simp [ball]) continuous_norm.continuousOn
\end{lstlisting}
\end{tcolorbox}

\subsubsection{Evaluation Problems}
Evaluation problems require the explicit synthesis of a solution term rather than a mere existence proof. These tasks are categorized into two subclasses: numerical computation, where the problem parameters are concrete numbers, and symbolic derivation, where the solution must be expressed in terms of abstract variables.
Below is a representative example of a numerical computation problem involving quadratic programming (QP).

\textbf{Informal Statement.} Find an optimal point of the following optimization problem.
\[
\begin{array}{ll}
\text{minimize} & (1/2)x^T P x + q^T x + r \\
\text{subject to} & -1 \leq x_i \leq 1, \quad i = 1, 2, 3,
\end{array}
\]

where  
\[
P = 
\begin{bmatrix}
13 & 12 & -2 \\
12 & 17 & 6 \\
-2 & 6 & 12
\end{bmatrix}, \quad q = 
\begin{bmatrix}
-22.0 \\
-14.5 \\
13.0
\end{bmatrix}, \quad r = 1.
\]

The example above is a typical numerical computation problem. In the problem, we provide several numerical matrix and vectors, and the task is to construct an optimal point numerically. To complete the lean content, it is required to figure out the result and fill it in the first \textit{sorry}, and then prove the optimality of the result in the theorem.

\textbf{Formal Statement.}
\begin{tcolorbox}[
  colback=gray!10,
  colframe=black,
  boxrule=0.5pt,
  arc=1mm,
  left=2mm, right=2mm,
  top=0.2mm, bottom=0.2mm
]
\begin{lstlisting}[language=Lean]
def answer : Fin 3 → ℝ := sorry

theorem num_3_S_E : let x := answer
  let P : Matrix (Fin 3) (Fin 3) ℝ:= !![13, 12, -2; 12, 17, 6; -2, 6, 12]
  let q : Fin 3 → ℝ := ![-22, -14.5, 13]
  let r := 1
  IsMinOn (fun y => (1 / 2) * (y ⬝ᵥ P *ᵥ y) + (q ⬝ᵥ y) + r) {x | ∀ i : Fin 3, -1 ≤ x i ∧ x i ≤ 1} x
  ∧ ∀ i : Fin 3, -1 ≤ x i ∧ x i ≤ 1:= by sorry
\end{lstlisting}
\end{tcolorbox}

\textbf{Formalization Challenge.} This example represents a typical numerical computation task. The problem inputs (matrix $P$ and vectors $q, r$) are provided as concrete data structures. The agent must compute the specific numerical vector $x^*$ that minimizes the objective function and instantiate it in the definition block (resolving the first \texttt{sorry}). Meanwhile, the agent must subsequently provide a rigorous proof in the theorem block showing that this specific candidate $x^*$ satisfies the optimality conditions under the given constraints.
Unlike numerical problems, symbolic derivation tasks require the agent to construct a general closed-form expression in terms of abstract parameters.

\textbf{Informal Statement.} Derive the analytical solution for the unconstrained quadratic minimization problem:

$$\min_{x \in \mathbb{R}^n} \quad f(x) = \frac{1}{2} x^T A x + b^T x,$$
where $A \in \mathbb{R}^{n \times n}$ is a symmetric positive definite matrix ($A \succ 0$) and $b \in \mathbb{R}^n$ is an arbitrary constant vector.

\textbf{Formal Statement.}
\begin{tcolorbox}[
  colback=gray!10,
  colframe=black,
  boxrule=0.5pt,
  arc=1mm,
  left=2mm, right=2mm,
  top=0.2mm, bottom=0.2mm
]
\begin{lstlisting}[language=Lean]
variable (n : Nat) (A : Matrix (Fin n) (Fin n) ℝ) (b : Fin 3 → ℝ)

def answer (n : Nat) (A : Matrix (Fin n) (Fin n) ℝ) (b : Fin 3 → ℝ) : Fin n → ℝ := sorry

theorem num_3_S_E : let x := answer n A b
  IsMinOn (fun y => (1 / 2) * (y ⬝ᵥ A *ᵥ y) + (b ⬝ᵥ y) + c) .univ x := by sorry


\end{lstlisting}
\end{tcolorbox}

\textbf{Formalization Challenge.}
This problem tests the agent's ability to perform symbolic algebraic manipulation rather than arithmetic calculation. The agent must define a function that takes the parameters $A$ and $b$ as arguments and returns the term $-A^{-1}b$ (or an equivalent representation involving the linear solve operation). The definition must be structurally valid for any $A$ and $b$ satisfying the type constraints. The subsequent theorem requires proving that this constructed term is indeed the unique global minimizer. This involves invoking properties of positive definite matrices and gradients (e.g., showing $\nabla f(x^*) = 0$ or completing the square) rather than checking numerical inequalities.

\subsubsection{Algorithm Design}

\textbf{Informal Statement.} Suppose $L \in \mathbb{R}^{n \times n}$ are lower triangular and $B \in \mathbb{R}^{n \times n}$. Give an algorithm for computing $X \in \mathbb{R}^{n \times n}$ so that $L X=B$.

\textbf{Formal Statement.}
\begin{tcolorbox}[
  colback=gray!10,
  colframe=black,
  boxrule=0.5pt,
  arc=1mm,
  left=2mm, right=2mm,
  top=0.2mm, bottom=0.2mm
]
\begin{lstlisting}[language=Lean]
open Matrix WithLp

variable {n : ℕ} {L : Matrix (Fin n) (Fin n) ℝ} (B : Matrix (Fin n) (Fin n) ℝ)

variable (hL : L.BlockTriangular OrderDual.toDual)

def algorithm (hL : L.BlockTriangular OrderDual.toDual) (B : Matrix (Fin n) (Fin n) ℝ) : ℕ → Matrix (Fin n) (Fin n) ℝ := fun k => match k with
  | 0 => 0
  | k + 1 => sorry

theorem algorithm_prop (B : Matrix (Fin n) (Fin n) ℝ) :
    L * algorithm hL B n = B := by
  sorry
\end{lstlisting}
\end{tcolorbox}

\textbf{Formalization Challenge.}
This problem category evaluates the agent's ability to synthesize algorithmic logic rather than static solutions. Unlike evaluation tasks where the output is a closed-form term, here the target is a state-transition rule.
To resolve the first \texttt{sorry}, the agent must explicitly implement a forward substitution algorithm leveraging the lower-triangular structure of. This requires translating the standard imperative loop such as computing the -th component based on previous results into Lean's functional recursive style, effectively defining the step-by-step evolution of the matrix .
The verification phase (the second \texttt{sorry}) demands a proof of algorithmic correctness. The agent must demonstrate via induction that the constructed sequence terminates at the exact solution after steps. This tests the capability to model control flow and establish loop invariants, bridging the gap between numerical algorithm design and formal logic.

\subsubsection{Representation Transformation}
We adhere to the theoretical framework established in Section \ref{subsec:rep_trans} and leverage the foundational class definitions provided in Appendix~\ref{subsec:class_transformation}, which implement these concepts in Lean. This category evaluates an agent's ability to recognize underlying mathematical structures and map them to tractable canonical forms.
Below, we present a representative example involving the reduction of a matrix approximation problem to a linear program (LP).

\textbf{Informal Statement.} We are given $k + 1$ matrices $A_0, \ldots, A_k \in \mathbb{R}^{m \times n}$, and need to find $x \in \mathbb{R}^k$ that minimizes
\[
\|A_0 + x_1 A_1 + \cdots + x_k A_k\|_{\infty}.
\]
Express this problem as a linear program.

\textbf{Formal Statement.}
\begin{tcolorbox}[
  colback=gray!10,
  colframe=black,
  boxrule=0.5pt,
  arc=1mm,
  left=2mm, right=2mm,
  top=0.2mm, bottom=0.2mm
]
\begin{lstlisting}[language=Lean]
open scoped Matrix.Norms.Operator

variable (m n k : ℕ) (A : Fin k → Matrix (Fin m) (Fin n) ℝ)

noncomputable def P : OptProblem where
  n_var := k
  n_eqs := 0
  eq_constraints := 0
  n_ieqs := 0
  ieq_constraints := 0
  objective := fun x => ∥∑ i, x i • (A i)∥

def answer (m n k : ℕ) (A : Fin k → Matrix (Fin m) (Fin n) ℝ) : LP  := sorry

theorem num_7_T_H :
  let Q := answer m n k A
  let P := P m n k A
  equivalence P.toOriginalProblem Q.toOriginalProblem := by sorry

\end{lstlisting}
\end{tcolorbox}

\textbf{Formalization Challenge.}
This problem introduces a distinct layer of complexity compared to direct calculation. To resolve the first \texttt{sorry}, the agent must explicitly construct the linear programming form. This involves a non-trivial modeling step: understanding that minimizing an $L_\infty$ norm is equivalent to minimizing a scalar slack variable $t$ subject to linear constraints ($-t \leq (A(x))_{ij} \leq t$).
Consequently, the agent must correctly define a new decision space that includes the auxiliary slack variable (mapping from $\mathbb{R}^k$ to $\mathbb{R}^{k+1}$). The agent also needs to populate the fields of the LP structure with the correct block-matrix representations derived from the input matrices $A_i$. Furthermore, the verification theorem requires unfolding the custom LP definitions found in the Appendix. This demands that the LLM not only process the immediate local context but also retrieve and comprehend the global type definitions to construct a valid proof of equivalence. This challenge the agent's capability on the multi-modal interplay between natural language mathematical intent and specific code library constraints) 

\section{Details and Analysis on AMBER Benchmark}
\label{sec:benchmark_analysis}

\begin{table*}[t]
    \centering
    \caption{Detailed statistics of the benchmark dataset across different domains.}
    \label{tab:dataset_stats}
    \begin{tabular}{lccccccc}
        \toprule
        & \multicolumn{2}{c}{\textbf{Difficulty}} & \multicolumn{4}{c}{\textbf{Problem Category}} & \\
        \cmidrule(lr){2-3} \cmidrule(lr){4-7}
        \textbf{Domain} & \textbf{Easy} & \textbf{Hard} & \textbf{Proof} & \textbf{Evaluation} & \textbf{Algorithm} & \textbf{Transformation} & \textbf{Total} \\
        \midrule
        Optimization & 24 & 26 & 18 & 12 & 1 & 19 & 50 \\
        Convex Analysis & 26 & 24 & 36 & 13 & 1 & 0 & 50 \\
        Numerical Algebra & 25 & 25 & 19 & 12 & 19 & 0 & 50 \\
        High-dim Probability & 25 & 25 & 45 & 5 & 0 & 0 & 50 \\
        \midrule
        \textbf{Total} & 100 & 100 & 118 & 42 & 21 & 19 & \textbf{200} \\
        \bottomrule
    \end{tabular}
\end{table*}

\subsection{Term Analysis}
\label{subsec:term_analysis}

To construct a benchmark that evaluates deep mathematical reasoning, mere text scraping is insufficient. Unlike natural language, formal mathematics in Lean possesses a rigorous dependency structure where every symbol is unambiguously defined. We developed a custom static analysis engine to extract these semantic dependencies, ensuring that each theorem in our dataset is accompanied by the precise definitions of the mathematical concepts it employs.

\subsubsection{Environment Reconstruction and AST Traversal}
The first stage of our pipeline involves reconstructing the compilation environment for each source file. We utilize Lean's compiler infrastructure to elaborate the source code, transforming raw text into fully typed abstract syntax trees (ASTs). This step is crucial because it resolves imports and namespaces, allowing us to distinguish between effectively identical names (e.g., \texttt{Matrix.trace} vs. \texttt{LinearMap.trace}) based on their precise identity in the global environment.

For each target theorem $T$, we extract its statement (type signature) and perform a recursive traversal of its expression tree. Let $E_T$ denote the expression tree of theorem $T$. We collect the set of all unique constants $C_T = \{c \mid c \in \text{leaves}(E_T)\}$ appearing in the statement. This process captures not just the top-level concepts but also nested dependencies used in arguments or return types.

\subsubsection{Semantic Stopword Filtering}
A raw extraction of constants yields a mixture of high-level mathematical concepts and low-level logical infrastructure. To focus on domain-specific reasoning, we implement a semantic stopword filter.

We partition the set of constants $C_T$ into two subsets: $C_{\text{infra}}$ and $C_{\text{domain}}$.

\noindent
\textbf{Infrastructure Terms ($C_{\text{infra}}$).} 
These include logical connectives ($\wedge, \vee, \to$), equality types (\texttt{Eq}), basic data types (\texttt{Nat}, \texttt{String}), and internal compiler auxiliary definitions. These are treated as stop words similar to NLP preprocessing, as they provide structural glue rather than mathematical content.

\medskip

\noindent
\textbf{Domain Concepts ($C_{\text{domain}}$).} 
These are the definitions, structures, and classes relevant to the mathematical subfield (e.g., \texttt{Convex}, \texttt{Measure}, \texttt{Eigenvalue}).

\medskip

Our analyzer filters out $C_{\text{infra}}$, retaining only the domain-specific concepts that a model must understand to prove the theorem.

This pipeline results in a structured representation for each theorem $T$, consisting of the tuple $(S_T, \mathcal{D}_T)$, where $S_T$ is the formal statement and $\mathcal{D}_T$ is the set of resolved dependencies, each containing its formal definition, source code location, and natural language documentation. The listing below presents a concrete example of this JSON structure for a specific theorem, showcasing the resolved dependencies and their associated metadata.

\begin{tcolorbox}[
  enhanced,
  breakable,
  colback=gray!10,
  colframe=black,
  boxrule=0.5pt,
  arc=1mm,
  left=2mm, right=2mm,
  top=0.2mm, bottom=0.2mm
]
\begin{lstlisting}[language=json]
{
  “theoremName": “isUnit_one_sub_skew_symmetric",
  “statement": “∀ {n : ℕ} [NeZero n] (S : Matrix (Fin n) (Fin n) ℝ), ...",
  “filePath": “Benchmarks/NAlg/num_2_P_E.lean",
  “dependencies": [
    {
      “name": “Matrix.one",
      “module": “Mathlib.Data.Matrix.Diagonal",
      “kind": “definition",
      “docString": null,
      “depth": 1,
      “definition": “{n : Type u_3} → ... → One (Matrix n n α)",
      “defRange": {“startLine": 199, “endLine": 200}
    },
    {
      “name": “Matrix.transpose",
      “module": “Mathlib.LinearAlgebra.Matrix.Defs",
      “kind": “definition",
      “docString": “The transpose of a matrix.",
      “depth": 1,
      “definition": “{m : Type u} ... → Matrix m n α → Matrix n m α",
      “defRange": {“startLine": 133, “endLine": 135}
    },
    {
      “name": “IsUnit",
      “module": “Mathlib.Algebra.Group.Units.Defs",
      “kind": “definition",
      “docString": “An element `a : M` of a `Monoid` is a unit if...",
      “depth": 1,
      “definition": “{M : Type u} → [Monoid M] → M → Prop",
      “defRange": {“startLine": 362, “endLine": 369}
    },
    ... 
  ]
}
\end{lstlisting}
\end{tcolorbox}

\subsection{Domain Statistics and Benchmark Fingerprinting}
\label{subsec:domain_statistics}

Leveraging the dependency extraction pipeline described above, we performed a quantitative analysis of the mathematical domains covered by our benchmark compared to existing datasets, specifically FATE (focused on algebra) and MiniF2F (focused on high school competitions).
Since Mathlib's directory structure strictly follows mathematical hierarchies, mapping the filtered domain constants $C_{\text{domain}}$ back to their defining modules provides a high-resolution fingerprint of the dataset's domain focus.

We aggregated the frequency of imported modules for all theorems in each dataset, excluding common utility files. The results, summarized in Table~\ref{tab:domain_fingerprint}, reveal distinct domain signatures that align with the intended scope of each benchmark.

\begin{itemize}
    \item \textbf{FATE (Algebraic Focus):} As expected for an algebra-centric benchmark, the dependencies are heavily concentrated in abstract algebraic structures. Prominent modules include \texttt{Mathlib.Algebra.Lie}, \texttt{Mathlib.Algebra.Polynomial}, \texttt{Mathlib.RingTheory.Ideal} and \texttt{Mathlib.Algebra.EuclideanDomain}. This confirms its specialized utility for evaluating reasoning in commutative and non-commutative algebra.

    \item \textbf{MiniF2F (Olympiad Focus):} The distribution reflects the nature of pre-university mathematics competitions, which prioritize elementary number theory, combinatorics, and discrete mathematics. The dominant modules are \texttt{Mathlib.Data.Nat} (natural numbers), \texttt{Mathlib.Data.Finset} (finite sets), and \texttt{Mathlib.Data.Nat.Cast}, indicating a focus on arithmetic manipulation and finite structures rather than continuous analysis.

    \item \textbf{ProofNet (Undergraduate Analysis Focus):} The dependency profile indicates a strong emphasis on the standard undergraduate analysis and topology curriculum. Leading imports include \texttt{Mathlib.Analysis.Normed.Group}, \texttt{Mathlib.Topology.Defs}, and \texttt{Mathlib.Topology.MetricSpace}. This fingerprint reflects tasks centered on metric spaces, continuity, and fundamental properties of normed vector spaces.
    
    \item \textbf{AMBER (Applied Mathematics):} In contrast, our dataset exhibits a unique footprint characterized by continuous optimization and high-dimensional analysis. The top dependencies include \texttt{Mathlib.LinearAlgebra.Matrix} and \texttt{Mathlib.Analysis.InnerProductSpace}, which are foundational for numerical linear algebra and convex optimization. Furthermore, the significant presence of \texttt{Mathlib.MeasureTheory.MeasurableSpace} reflects the inclusion of high-dimensional probability tasks. This distribution empirically verifies that our AMBER benchmark successfully targets the gap in applied computational mathematics.
\end{itemize}

\begin{table}[t]
\centering
\caption{Top-3 Mathlib module dependencies across different benchmarks. The statistics highlight distinct specializations: FATE (Abstract Algebra), MiniF2F (Discrete Math), ProofNet (Topology/Analysis), and AMBER (Numerical/Applied Analysis).}
\label{tab:domain_fingerprint}
\begin{tabular}{lll}
\toprule
\textbf{Benchmark} & \textbf{Top Modules (Domain Signature)} & \textbf{Mathematical Focus} \\ 
\midrule
\multirow{3}{*}{FATE} 
 & Mathlib.Algebra.Category.ModuleCat & Module Categories \\
 & Mathlib.RingTheory.Ideal.Quotient & Quotient Rings \\
 & Mathlib.FieldTheory.IntermediateField & Field Theory \\ 
\midrule
\multirow{3}{*}{MiniF2F} 
 & Mathlib.Data.Nat & Number Theory \\
 & Mathlib.Data.Finset & Combinatorics \\
 & Mathlib.Data.Nat.Cast & Arithmetic \\ 
\midrule
\multirow{3}{*}{ProofNet} 
 & Mathlib.Analysis.Normed.Group & Normed Spaces \\
 & Mathlib.Topology.Defs & General Topology \\
 & Mathlib.Topology.MetricSpace & Metric Spaces \\ 
\midrule
\multirow{4}{*}{\textbf{AMBER}} 
 & Mathlib.LinearAlgebra.Matrix & Numerical Linear Algebra \\
 & Mathlib.Analysis.InnerProductSpace & Optimization \\
 & Mathlib.Analysis.Convex & Convex Analysis \\
 & Mathlib.MeasureTheory.MeasurableSpace & Probability \\ 
\bottomrule
\end{tabular}
\end{table}

\subsection{Difficulty Analysis}
\label{difficulty_analysis}
To ensure the benchmark serves as a robust diagnostic tool for varying levels of mathematical capability, we enforced a rigorous difficulty stratification during the collection process.
We classify problem difficulty based on the academic maturity required to solve the source material.
Easy problems are derived from standard undergraduate course exercises, typically requiring direct application of definitions or single-step inferences.
Hard problems are sourced from graduate-level textbooks and PhD qualifying examinations, necessitating multi-step reasoning, creative auxiliary constructions, or deep structural insights.
Table~\ref{tab:dataset_stats} presents the distribution of difficulty levels across domains.
A distinct feature of our benchmark is the strict balance achieved: the dataset contains an exact split of 100 Easy and 100 Hard problems.
Crucially, this symmetry is maintained not only globally but also locally within each subfield.
This uniform stratification prevents the evaluation from being skewed by domain-specific difficulty spikes and allows researchers to assess whether an agent's performance degradation on hard tasks is consistent across different mathematical subjects.
Furthermore, the difficulty classification adapts to the specific paradigms introduced in Section~\ref{formalization_am}.
For theorem proving tasks, difficulty correlates with the logical depth and the sparsity of the required search path.
In contrast, for construction-verification tasks (evaluation, algorithm, transformation), difficulty is governed by the structural complexity of the target object.
For instance, an easy construction task might involve finding a scalar root, whereas a hard task requires designing an iterative matrix factorization algorithm or constructing a complex semidefinite relaxation.
This multi-dimensional difficulty structure ensures that the benchmark can effectively discriminate between models that merely memorize syntactic patterns and those capable of genuine mathematical synthesis.

\subsection{LLM Verification Details}
\label{LLM_Verification_Details}
The exact prompt employed for the preliminary semantic consistency check is provided below. 

\textbf{System Prompt.}
\begin{tcolorbox}[
  colback=gray!10,
  colframe=black,
  boxrule=0.5pt,
  arc=1mm,
  left=2mm, right=2mm,
  top=2mm, bottom=2mm
]
You are an expert Mathematician and Lean 4 Formalization Engineer. Your task is to verify the strict semantic consistency between a natural language mathematical problem and its formalized Lean code. You must detect any logical omissions, semantic discrepancies, or misalignments in variable definitions, constraints, or the final goal.
\end{tcolorbox}

\textbf{User Prompt.}
\begin{tcolorbox}[
  enhanced,
  breakable,
  colback=gray!10,
  colframe=black,
  boxrule=0.5pt,
  arc=1mm,
  left=2mm, right=2mm,
  top=2mm, bottom=2mm
]
Please review the following pair:

1. Natural Language Problem: \{natural\_language\_problem\}

ORIGINAL\_Natural\_Language\_Problem\_START

\{natural\_language\_problem\}

ORIGINAL\_Natural\_Language\_Problem\_END

2. Lean Formalization: 

ORIGINAL\_LEAN\_FILE\_START

\{lean\_source\}

ORIGINAL\_LEAN\_FILE\_END

Validation Instructions:

    Variable Mapping: Verify that every variable in the text exists in the code with the correct type (e.g., real vs. integer, non-negative constraints).

    Hypothesis Check: Ensure all conditions (e.g., "distinct," "prime," "x>0") are explicitly present in the Lean hypotheses.

    Goal Alignment: Check if the proof goal in Lean strictly matches the question asked in the text.

    Mathlib Usage: Verify that standard concepts (primes, factorials, geometry definitions) use standard Mathlib definitions and not incorrect custom ones.

Output: If the code is correct, output: [CONSISTENT]. If there are errors, output: [MISMATCH] followed by a concise explanation of the missing or incorrect logic.
\end{tcolorbox}

\section{Experiment Details}
\label{sec:experiment_details}
\subsection{Baseline Setup}
\label{subsec:baseline_setup}

To establish a comprehensive evaluation landscape, we select a diverse set of models ranging from state-of-the-art general reasoning engines to specialized open-source theorem provers.We access proprietary models via their official APIs, specifically utilizing DeepSeek-V3.2-Thinking and GPT-5.1. Complementing these, we locally deploy prominent open-source models fine-tuned for formal mathematics, including Kimina-Prover and Goedel-Prover, to assess the impact of domain-specific training. To ensure reproducibility and representativeness of standard usage, we adhere to the default inference parameters, including temperature and top-$p$, recommended by the respective official documentation for all models. The maximum generation length is uniformly set to 32,768 tokens to accommodate the verbose nature of formal proofs and constructive definitions. For reasoning-enhanced models that support adjustable compute allocation, we explicitly configure the reasoning effort parameter to medium.

\subsection{Prompts}
\label{subsec:prompts}
We provide our detailed prompts used for evaluating existing models. 

\textbf{System Prompt.}
\begin{tcolorbox}[
  colback=gray!10,
  colframe=black,
  boxrule=0.5pt,
  arc=1mm,
  left=2mm, right=2mm,
  top=2mm, bottom=2mm
]
You are a Lean 4 theorem-proving assistant.

Task:

- You will receive the content of a Lean 4 file for a math problem.

- You must produce a COMPLETE Lean 4 source file that can be compiled in the given project.

Strict output requirements:

- Give the complete proof of the statement. DO NOT contain sorry in your output

- DO NOT change anything in the file except for the sorry.

- Output ONLY the Lean 4 source code of the final file.

- DO NOT add any explanations, comments about what you are doing, or natural language text outside of the code.

- DO NOT wrap the code in Markdown fences.

- DO NOT change the definitions and classes in the source file.

- The output must be a self-contained Lean 4 file that can be saved as a .lean file and compiled directly.

- If you receive a problem about solving a problem, you must produce a solution that is correct and complete. Do NOT use the problem itself to answer it.

- If you receive a problem about designing an algorithm, you must produce a complete algorithm that is correct and complete.You need to consider whether the algorithm can be realized, rather than only considering the correctness of the algorithm.
\end{tcolorbox}

\textbf{User Prompt.}
\begin{tcolorbox}[
  enhanced,
  breakable,
  colback=gray!10,
  colframe=black,
  boxrule=0.5pt,
  arc=1mm,
  left=2mm, right=2mm,
  top=2mm, bottom=2mm
]
Below is the content of a Lean 4 file for a math problem.

ORIGINAL\_LEAN\_FILE\_START

\{lean\_source\}

ORIGINAL\_LEAN\_FILE\_END

Please produce the final complete Lean 4 source file.

Your task is to replace each occurrence of the token `sorry` with correct Lean code.

You MUST NOT modify any other part of the file: all lines, imports, comments, names, and formatting outside `sorry` must remain exactly the same.

Do not introduce any new `sorry` or `submit`

Output only raw Lean code, no Markdown fences, no explanations.
\end{tcolorbox}

\subsection{Unbiased Estimation for Pass@k Results}
\label{subsec:unbiased}
We report the Pass@k metric using the unbiased estimator~\cite{chen2021evaluating}. Evaluating Pass@k by simply generating $k$ samples and checking if at least one is correct results in high variance. To address this, we generate a larger pool of $n$ samples ($n \ge k$) for each problem and estimate the probability that at least one sample is correct if we were to draw $k$ samples from this pool without replacement.Let $n$ be the total number of samples generated for a problem, and let $c$ be the number of correct samples among them. The unbiased estimator for Pass@k is calculated as:\begin{equation}\text{Pass}@k := \mathbb{E}\left[1 - \frac{\binom{n-c}{k}}{\binom{n}{k}}\right]\end{equation}where $\binom{n}{k}$ denotes the binomial coefficient. In our experiments, we set $k=1, 2, 4, 8, 16$ and the total sample size $n=16$. When $n=k$, this formula simplifies to the empirical accuracy (1 if $c > 0$, 0 otherwise). For cases where $n-c < k$, the term $\binom{n-c}{k}$ is defined as 0, resulting in a Pass@k of 1.

\label{appendix:pass_k_stats}

Table \ref{tab:pass_k_all_models} presents the detailed Pass@k statistics ($k=\{1, 2, 4, 8, 16\}$) for all evaluated models. The metrics are calculated using the unbiased estimator described above, based on $n=16$ samples per problem.

\begin{table}[h]
\centering
\caption{Pass@k ($k=1, 2, 4, 8, 16$) statistics for all models on the full benchmark. The values represent the percentage of problems solved given $k$ budget.}
\label{tab:pass_k_all_models}
\begin{tabular}{lccccc}
\toprule
\textbf{Model} & \textbf{Pass@1} & \textbf{Pass@2} & \textbf{Pass@4} & \textbf{Pass@8} & \textbf{Pass@16} \\
\midrule
\multicolumn{6}{l}{\textit{General-Purpose Reasoning Models}} \\
DeepSeek-V3.2-Thinking & 1.68\% & 2.71\% & 4.04\% & 4.86\% & 6.00\% \\
GPT-5.1               & 1.40\% & 2.27\% & 3.27\% & 4.20\% & 4.50\% \\
Gemini-3.0 Pro         & 2.62\% & 3.52\% & 4.46\% & 4.41\% & 6.00\% \\
\midrule
\multicolumn{6}{l}{\textit{Specialized Theorem Provers}} \\
Goedel Prover-32B      & 0.97\% & 1.70\% & 2.66\% & 3.50\% & 4.00\% \\
Kimina Prover-72B      & 1.03\% & 1.65\% & 2.51\% & 3.50\% & 4.00\% \\
\bottomrule
\end{tabular}
\end{table}

\subsection{Ablation Study on Prompts}
\label{subsec:ablation}

To rigorously evaluate the impact of prompt engineering on model performance, we conducted a targeted ablation study on a subset of 20 specialized problems selected from the optimization and convex analysis domains. This study aimed to determine whether the structural constraints of our construction-verification framework are necessary for applied mathematical reasoning or if standard formalization prompts suffice. We compared three distinct prompting strategies provided to the models.
\begin{itemize}
    \item \textbf{Construction-Verification.} The model is explicitly required to first define a computable solution term via a def block and then prove its correctness within a theorem block.
    \item \textbf{Standard Code Completion.} The model receives a Lean file with \texttt{sorry} placeholders but no explicit structural guidance or emphasis on the two-stage workflow.
    \item \textbf{Traditional Proof Only.} The model is tasked solely with proving the provided statement without the requirement to define an explicit solution object.
\end{itemize}

As shown in Table~\ref{tab:ablation_prompts}, the construction-verification prompt yielded the highest success rate, while the other two strategies significantly misled the models. Our analysis indicates that traditional proof-only prompts often trigger a 40\% decrease in validity compared to our structured approach. Without the explicit requirement for construction, models frequently resort to non-constructive tactics such as \texttt{Classical.choice} or apply is\_compact.exists\_min. While these methods logically prove the existence theorem, they provide no computational value, failing the core objective of applied mathematics. Furthermore, standard completion prompts without structural guardrails often lead to ``tactical overfitting", where the model ignores the def block entirely and attempts to discharge the goal using proof-search tactics alone. These results confirm that explicit construction-verification represents a distinct paradigm that necessitates a tailored formalization interface to prevent models from bypassing the rigorous derivation of solutions.

\section{Details on Instructions Following Analysis}
\label{sec:details_instruction}

A counter-intuitive finding in our experiments is the inverse correlation between a model's specialization in formal proving and its performance on applied mathematics tasks. As shown in Table~\ref{tab:results}, specialized models exhibit performance comparable to, or even exceeding, general-purpose LLMs on standard theorem proving tasks. However, their capabilities degrade precipitously on algorithm design and representation transformation tasks. We identify this phenomenon as the ``tactical overfitting" effect. Specialized models are fine-tuned extensively on state-tactic pairs, which maximizes their ability to close logical goals, but also appears to compromise their instruction-following capabilities in two critical ways relevant to applied mathematics.

\textbf{Rigidity in Problem Structure.} Applied mathematics problems often inherently  a complex structure as described in Section \ref{formalization_am}, where the agent must first synthesize a constructive definition before proving its properties. General-purpose models, retaining strong instruction-following abilities from their base pre-training, respect this structural requirement. In contrast, specialized models frequently ignore the instruction to define data structures, attempting instead to solve the entire problem within a single \texttt{theorem} block. This ``hammer-and-nail" bias of treating every prompt as a pure theorem proving task leads to disqualification in constructive benchmarks.

\textbf{Loss of Applied Mathematics Intuition.} General-purpose models demonstrate a superior ability to translate informal mathematical concepts into formal specifications. For instance, when asked to formalize a relaxation to an SDP, general models often correctly infer the necessary matrix inequalities. Specialized models, despite being stronger at manipulating algebraic expressions once formalized, struggle to generate the initial formal definition from natural language descriptions. They effectively lack the common sense to bridge the gap between informal intent and formal syntax, a capability likely eroded during intensive fine-tuning on pure proof corpora. Consequently, while specialized provers remain superior for distinct logical verification steps, they lack the flexibility required for the end-to-end workflow of an applied mathematician, which heavily relies on structural synthesis and cross-domain modeling.

Our investigation into model behavior reveals a significant discrepancy in how different architectures handle the multi-stage requirements of applied mathematical formalization. While specialized provers excel in isolated logical deduction, they frequently fail the end-to-end constructive workflow.

\subsection{The Tactical Overfitting Phenomenon} 
We characterize the performance degradation in specialized models as ``tactical overfitting". This effect is observed when models fine-tuned exclusively on proof-state-tactic pairs develop a bias toward closing proof goals at the expense of structural synthesis.

A salient example of the ``tactical overfitting" effect can be observed in the behavior of specialized models like Kimina-Prover during complex, structured reasoning tasks. In the following instance, the model attempted to utilize the \texttt{intro} tactic to invoke preceding variables that had been established within a \texttt{let} statement, a ubiquitous procedure in traditional declarative proofs but a structural error in this specific context. This mistake highlights a critical lack of situational awareness; because the model has been fine-tuned so intensively on state-tactic pairs, it defaults to high-frequency proof-search routines without correctly recognizing the formal environment or the existing variable scope.

Such failures represent a broader ``hammer-and-nail" bias where specialized provers treat every prompt as a pure theorem-proving task, effectively eroding the common sense intuition necessary to bridge the gap between informal mathematical intent and formal syntax. Like a master carpenter who only knows how to use a hammer, the model treats every structural challenge as a nail to be driven in with a proof tactic, ultimately lacking the flexibility required for the end-to-end workflows of an applied mathematician. Consequently, while specialized provers remain superior for distinct deductive steps, they fail to respect the multi-stage constructive requirements that general-purpose reasoning models navigate with much higher fidelity.

\begin{tcolorbox}[
  colback=gray!10,
  colframe=black,
  boxrule=0.5pt,
  arc=1mm,
  left=2mm, right=2mm,
  top=0.2mm, bottom=0.2mm,
]
\begin{lstlisting}[language=Lean, escapeinside={(*@}{@*)}]
variable (n m : ℕ) [NeZero n] [NeZero m] (A : Matrix (Fin m) (Fin n) ℝ) (b : Fin m → ℝ) (x : Fin n → ℝ)

noncomputable def P : OptProblem where
  n_var := n
  n_eqs := 0
  eq_constraints := 0
  n_ieqs := m
  ieq_constraints := fun x => A *ᵥ x - b
  objective := fun x => - ∑ i, log ((b - A *ᵥ x) i)

def Q : DualProblem (P n m A b) := 
  { 
    dual_objective := fun (lam : Fin 0 → ℝ) (mu : Fin m → ℝ) => 
      if h : ∀ i : Fin m, mu i ≥ 0 then 
        EReal.ofReal (- ∑ i : Fin m, (log (mu i) + 1))
      else 
        ⊥, 
    dual_domain := Set.filter (fun (p : (Fin 0 → ℝ) × (Fin m → ℝ)) => 
      let (lam, mu) := p
      (∀ i : Fin m, mu i ≥ 0)
    ) (Set.univ : Set ((Fin 0 → ℝ) × (Fin m → ℝ))), 
    h_objective := by 
      (*@\redwave{\texttt{intro lam mu}}@*) 
      simp [P, EReal, EReal.ofReal]
      all_goals try { 
        aesop 
      }
    , 
    h_domain := by
      ext ⟨lam, mu⟩
      simp [P, EReal, EReal.ofReal]
      all_goals try { 
        aesop
      }
  }
\end{lstlisting}
\end{tcolorbox}

\subsection{Loss of Applied Mathematics Intuition}

A critical requirement for applied mathematicians is the ability to recognize underlying mathematical structures and map them to tractable forms, such as deriving a dual problem from a primal formulation. Our experiments demonstrate that specialized theorem provers frequently lack the necessary modeling intuition to perform this translation accurately. While these models are proficient at algebraic manipulation once a goal is established, they struggle to bridge the gap between informal mathematical intent and formal syntax.

A representative failure is observed in the behavior of Kimina-Prover when tasked with deriving a dual problem. Instead of correctly constructing the Lagrangian and identifying the appropriate dual variables and constraints , the model attempted to produce a solution that was fundamentally misaligned with the problem's intent. Rather than evaluating the structural requirements of the dual space, the model defaults to generating correct-looking but logically incorrect definitions that mimic standard proof patterns.

In contrast, general-purpose models demonstrate a superior capability to infer necessary matrix inequalities and structural constraints from natural language descriptions. This case study underscores that the end-to-end workflow of applied mathematics, which relies heavily on structural synthesis and cross-domain modeling, remains a significant challenge for models over-optimized for logical verification alone.

\begin{tcolorbox}[
  colback=gray!10,
  colframe=black,
  boxrule=0.5pt,
  arc=1mm,
  left=2mm, right=2mm,
  top=0.2mm, bottom=0.2mm,
]
\begin{lstlisting}[language=Lean, escapeinside={(*@}{@*)}]
/-
Derive a dual problem for
\[
\begin{array}{ll}
\text{min} \quad &\sum_{i=1}^{N} \|y_i\|_2 + (1/2)\|x - x_0\|_2^2. \\
\text{subject to} & y_i = A_ix+b_i, i=1, 2, \cdots n
\end{array}
\]
The problem data are \( A_i \in \mathbb{R}^{m_i \times n},  b_i \in \mathbb{R}^{m_i} \), and  \( x_0 \in \mathbb{R}^n \).
-/
noncomputable def P : OptProblem where
    (*@\texttt{$\cdots$}@*)--omit detailed definition of P

def Q : DualProblem (P n m k A b x0) := by
  dual_objective := fun lam _ =>
    if h : ∀ i : Fin k, sqrt ((b i) ⬝ᵥ (b i)) ≤ 1 then
      EReal.ofReal (∑ i : Fin k, (b i) ⬝ᵥ (b i) + (A i *ᵥ x0) ⬝ᵥ (b i) - (1 / 2) * ((∑ i : Fin k, (A i : Matrix (Fin m) (Fin n) ℝ) ^* * (b i)) ⬝ᵥ (∑ i : Fin k, (A i : Matrix (Fin m) (Fin n) ℝ) ^* * (b i))) : ℝ)
      --Here the dual objective is wrong.
    else
      ⊥
  (*@\texttt{$\cdots$}@*)--omit detailed definition of Q


\end{lstlisting}
\end{tcolorbox}

\section{Details on Error Analysis}
\label{sec:details_error}
In this section, we provide comprehensive case studies to show the error information from the Lean 4 compiler.

\subsection{Hallucination of LLMs.}
LLMs frequently hallucinate non-existent theorems or definitions in formal reasoning settings. A representative example below in Lean 4 shows that the model attempts to complete a proof by invoking a non-existing theorem, \texttt{convexHull\_eq\_sInter}, confidently treating it as a valid result. This behavior suggests that the model is not reasoning within the formal system itself, but rather relying on memorized patterns from previously seen mathematical texts. Consequently, the apparent theorem-proving ability of current LLMs is unlikely to be genuinely transferable, as it fails to generalize to domains, libraries, or formalizations that were not explicitly encountered during training.

\begin{tcolorbox}[
  colback=gray!10,
  colframe=black,
  boxrule=0.5pt,
  arc=1mm,
  left=2mm, right=2mm,
  top=0.2mm, bottom=0.2mm,
]
\begin{lstlisting}[language=Lean, escapeinside={(*@}{@*)}]
--Show that the convex hull of a set \( S \) is
--the intersection ofall convex sets that contain \( S \).
theorem num_4_P_E (n : ℕ) [NeZero n] (C : Set (Fin n → ℝ)) :
    ∩ (S : Set (Fin n → ℝ)) (_ : C ⊆ S) (_ : Convex ℝ S) , S = convexHull ℝ C := by
    exact ((*@\redwave{\texttt{convexHull\_eq\_sInter}}@*) C).symm
\end{lstlisting}
\end{tcolorbox}

\subsection{Formalization Competence.} 

\begin{tcolorbox}[
  enhanced,
  breakable,
  colback=gray!10,
  colframe=black,
  boxrule=0.5pt,
  arc=1mm,
  left=2mm, right=2mm,
  top=0.2mm, bottom=0.2mm,
]
\begin{lstlisting}[language=Lean, escapeinside={(*@}{@*)}]
variable (n : ℕ) [NeZero n] (a : Fin n → ℝ) (b1 b2 : ℝ)

noncomputable def answer (n : ℕ) [NeZero n] (a : Fin n → ℝ) (b1 b2 : ℝ): ℝ :=
  |b1 - b2| / ∥a∥

lemma toLp2_norm_eq (v : Fin n → ℝ) : ∥toLp 2 v∥ = ∥v∥ :=
  (EuclideanSpace.equiv 2 (Fin n) ℝ).symm.norm_map v

theorem num_5_S_E (h : a ≠ 0):
  let A := {x | a ⬝ᵥ x = b1}
  let B := {x | a ⬝ᵥ x = b2}
  ∀ x y, A x → B y → ∥toLp 2 (x - y)∥ ≥ answer n a b1 b2 ∧
  ∃ x y, A x ∧ B y ∧ ∥toLp 2 (x - y)∥ = answer n a b1 b2 := by
  intro A B x y hx hy
  have hA : A x := hx
  have hB : B y := hy
  (*@\redwave{\texttt{simp\_rw [Set.mem\_setOf\_eq] at hA hB}}@*)
  have h1 : a ⬝ᵥ (x - y) = b1 - b2 := by
    rw [inner_sub_right, hA, hB, sub_sub_cancel]
  have hnorm_pos : ∥a∥ > 0 := norm_pos_iff.mpr h
  constructor
  · have h3 : |b1 - b2| ≤ ∥a∥ * ∥x - y∥ := by
      rw [← abs_inner_le_norm a (x - y)]
      exact congr_arg abs h1 ▸ le_refl _
    rw [toLp2_norm_eq, answer]
    exact (div_le_div_right hnorm_pos).mp h3
  · set x0 := (b1 / ∥a∥ ^ 2) • a
    set y0 := (b2 / ∥a∥ ^ 2) • a
    have hx0 : A x0 := by
      simp [A, Set.mem_setOf_eq, x0]
      rw [inner_smul_right, inner_self_eq_norm_sq, mul_div_cancel']
      exact pow_ne_zero 2 (norm_ne_zero_iff.mpr h)
    have hy0 : B y0 := by
      simp [B, Set.mem_setOf_eq, y0]
      rw [inner_smul_right, inner_self_eq_norm_sq, mul_div_cancel']
      exact pow_ne_zero 2 (norm_ne_zero_iff.mpr h)
    have hnorm : ∥toLp 2 (x0 - y0)∥ = answer n a b1 b2 := by
      rw [toLp2_norm_eq, answer]
      calc
        ∥x0 - y0∥ = ∥((b1 - b2) / ∥a∥ ^ 2) • a∥ := by simp [x0, y0, sub_smul, smul_sub]
        _ = |(b1 - b2) / ∥a∥ ^ 2| * ∥a∥ := norm_smul _ _
        _ = |b1 - b2| / ∥a∥ ^ 2 * ∥a∥ := by rw [abs_div, abs_of_pos (pow_pos hnorm_pos 2)]
        _ = |b1 - b2| / ∥a∥ := by field_simp [ne_of_gt hnorm_pos]; ring
    exact ⟨x0, y0, hx0, hy0, hnorm⟩
\end{lstlisting}
\end{tcolorbox}

A typical manifestation of insufficient formalization competence is illustrated by the error “simp made no progress” in the highlighted line. Here, the model applies the simplification tactic \texttt{simp\_rw} to the hypotheses \texttt{hA} and \texttt{hB}, implicitly assuming that this tactic would unfold the set membership and yield usable equalities. However, the tactic fails to simplify the goals in any meaningful way, indicating that no applicable rewrite rules were triggered. This reveals that the model incorrectly anticipates the effect of the \texttt{simp} tactic, treating it as a generic tool for progress rather than reasoning about its precise preconditions and rewrite behavior. Such errors suggest that the model lacks a concrete understanding of how tactics operate at the proof-state level, and instead selects tactics based on superficial associations rather than a correct grasp of their formal semantics.

\subsection{Incomplete Outputs.} 
A closely related manifestation of insufficient formalization competence is the model’s frequent reliance on special placeholders such as \texttt{sorry}, \texttt{submit}, or even newly introduced axioms to force incomplete proofs to pass the compiler’s checks. As illustrated in the example, the model successfully constructs the high-level mathematical objects and outlines the intended proof structure, but abandons the formal derivation when confronted with technically demanding arguments, such as verifying optimality conditions or completing inductive reasoning. Instead of resolving the remaining proof obligations, the model resorts to these escape mechanisms, which formally suppress unresolved goals rather than discharging them. This behavior indicates that the model treats the proof assistant as a permissive environment that can be coerced into accepting partial solutions, rather than as a system enforcing strict logical completeness. Consequently, while the generated code may appear syntactically valid, it fails to represent a genuine, machine-checked proof, further highlighting the gap between the model’s conceptual reasoning and its ability to carry out precise, low-level formal proofs.

\begin{tcolorbox}[
  colback=gray!10,
  colframe=black,
  boxrule=0.5pt,
  arc=1mm,
  left=2mm, right=2mm,
  top=0.2mm, bottom=0.2mm,
]
\begin{lstlisting}[language=Lean, escapeinside={(*@}{@*)}]
variable (n : ℕ) [NeZero n] (X : Matrix (Fin n) (Fin n) ℝ) (hX : X.PosDef)

def answer (n : ℕ) [NeZero n] (X : Matrix (Fin n) (Fin n) ℝ) (hX : X.PosDef) : Matrix (Fin n) (Fin n) ℝ :=
  let ⟨P, D, hD_diag, hPD, hPorth, hPorth'⟩ := hX.1.orthogonalDiagonalization
  let Λ : Matrix (Fin n) (Fin n) ℝ := Matrix.diagonal (fun i => (D i i + Real.sqrt ((D i i)^2 + 4)) / 2)
  P * Λ * Pᵀ

theorem num_43_S_H(n : ℕ) [NeZero n] (X : Matrix (Fin n) (Fin n) ℝ) (hX : X.PosDef) : let answer := answer n X hX
IsMinOn (fun U => -log (det U) + ((X - U) * (X - U)).trace / 2) .univ answer := by
  intro answer
  (*@\textcolor{red}{\uwave{\texttt{sorry}}}@*)
\end{lstlisting}
\end{tcolorbox}

\subsection{Formatting and Syntax.}
This category covers errors arising from improper output formatting or basic syntactic mistakes that prevent the compiler from correctly parsing the generated file. Typical cases include failing to wrap code in appropriate markdown or code blocks, producing malformed section or environment headers, and introducing missing or mismatched delimiters such as parentheses, brackets, or braces. In these instances, the generated content may be mathematically meaningful or even formally correct in isolation, yet it cannot be processed due to violations of the surface-level structural requirements imposed by the proof assistant or document system. Although these errors are comparatively shallow and do not reflect a lack of mathematical understanding, they nonetheless indicate that the model does not consistently internalize or enforce the rigid syntactic and formatting conventions required in formal verification settings. As a result, even minor deviations at the presentation level can lead to complete failure of compilation, underscoring the fragility of automated formalization pipelines with respect to low-level output correctness.

\section{Usage of LLM}
\label{sec:LLM_usage}
LLMs were evaluated as research subjects to assess their formalization capabilities within our proposed framework, while LLMs were also used to refine the manuscript's grammar and clarity.

\end{document}